\documentclass[fleqn,twoside]{article}%
\usepackage{amsmath}
\usepackage{amsfonts}
\usepackage{amssymb}
\usepackage{graphicx}%
\def\be{\begin{equation}}

\def\ee{\end{equation}}

\def\bi{\bibitem}

\begin{document}

\title{Analyzing Conserved Currents in F(R) theory of gravity}

\author{Nayem Sk$^1$, Manas Chakrabortty$^2$ and Abhik Kumar Sanyal$^3$}
\maketitle

\noindent

\begin{center}
$^1$Dept. of Physics, Saidabad Manindra Chandra Vidyapith, Murshidabad, India - 742103\\
$^2$Dept. of Physics, Bankura University, Bankura, India - 722155\\
$^3$Dept. of Physics, Jangipur College, Murshidabad, India - 742213\\
\end{center}

\footnotetext[1]{
Electronic address:\\

\noindent $^1$nayemsk1981@gmail.com\\
$^2$manas.chakrabortty001@gmail.com\\
\noindent $^3$sanyal\_ ak@yahoo.com\\}

\begin{abstract}

\noindent
F(R) theory of gravity is claimed to admit a host of conserved currents under the imposition of Noether symmetry following various techniques. However, for a constrained system such as gravity, Noether symmetry is not on-shell. As a result, the symmetries do not necessarily satisfy the field equations in general, constraints in particular, unless the generator is modified to incorporate the constraints. In the present manuscript, we apply the first theorem of Poisson to unveil the fact that not all the conserved currents appearing in the literature for F(R) theory of gravity, satisfy the field equations. We also provide a list of available forms of F(R) along with associated conserved currents, and construct a generalised action, which might address the cosmic puzzle, elegantly.\\

\end{abstract}

\section{Introduction}
$F(R)$ theory of gravity has drawn a lot of attention to the cosmologists in recent years as an alternative to the dark energy. It is said to have solved the cosmic puzzle elegantly and singlehandedly, unifying early inflation with late-time cosmic acceleration, and also it admits the Newtonian limit, so that the theory passes the solar test elegantly [see \cite{1a, 1b, 1c, 1d, 1e} for thorough reviews]. Other than reconstruction programme, in which one starts with available cosmological data and fit it with a suitable form of $F(R)$, Noether symmetry is usually imposed to find the form of $F(R)$ a-priori. To cast the action in canonical form so that Noether symmetry is applicable, initially Lagrange multiplier technique was adopted. Following this technique, $F(R) \propto R^{\frac{3}{2}}$, along with a conserved current $\frac{d}{dt}(a\sqrt R)$ in vacuum and matter dominated eras were found in the background of Robertson-Walker line-element, where `$a$' is the scale factor \cite{2a,2b,2c,2d,2e,2f,2g,2h,2i,2j}. Although such a form of $F(R) \propto R^{\frac{3}{2}}$ leads to power law inflation in the early universe and to some extent can explain late-time cosmic acceleration, it fails to provide a well-behaved radiation dominated era ($a \propto \sqrt t$) and early stage of matter dominance ($a\propto t^{2\over 3}$) \cite{2j}. On the contrary, it is worth mentioning that starting from $F(R) \propto R^2$ a-priori, modified scalar-tensor theory of gravity admits Noether symmetry, leading to power law inflation in vacuum \cite{3}. Further, in the case of nonminimally coupled scalar-tensor theory of gravity being modified by $F(R)$ term, a metric-independent general conserved current ${J^{;\mu}} = \left(\phi^{;\mu}\sqrt{3f'^2 + 2f{w\over \phi}}\right)$, (where, $f(\phi)$ and $w(\phi)$ are the coupling parameter and the Brans-Dicke parameter respectively and prime denotes derivative with respect to the scalar field), exists for traceless field (viz. in vacuum and radiation domains in the cosmological context) admitting arbitrary form of $F(R)$, provided the scalar potential $V(\phi)$ is proportional to the square of the coupling parameter $f(\phi)$ \cite{4a, 4b}. The same has also been found to be true for generalized Born-Infeld action, of-course with a different conserved current \cite{4c}. This fact signals that one can associate an indefinitely large number of symmetries in connection with non-minimally coupled scalar-tensor theory of gravity, being modified by $F(R)$ term, particularly in the vacuum and radiation dominated eras. However, on the contrary, following Lagrange multiplier technique, no admissible symmetry, other than $F(R) \propto R^{3\over 2}$, was found for $F(R)$ theory of gravity, following Noether symmetry analysis even by enlarging the configuration space under the addition of a scalar field or taking anisotropic models into account \cite{5}. Initially, this appeared to be an unresolved problem.\\

To understand the situation better and to get certain insights, Noether symmetry analysis was therefore performed in the scalar-tensor equivalent forms of $F(R)$ theory of gravity \cite{6}. Interestingly enough, while, Jordan's frame has been found to be identical to the Lagrange multiplier technique in the sense that it yields the same form of $F(R)$ along with identical conserved current, Einstein's frame appeared to be different, since it led to some new forms of $F(R)$ \cite{6}. In particular, it yields $F(R) \propto R^2$, together with a conserved current, $\Sigma = a^3 \dot R$ for $k = 0, \pm 1$ in the vacuum and also in the radiation dominated eras. In vacuum, it further leads to $F(R) \propto R^{-1}$, being associated with a conserved current, $\Sigma = R\dot a\sqrt{a}$, and also $F(R) \propto R^{7\over 5}$, carrying a conserved current, $\Sigma = \sqrt a {d\over dt}(a R^{2\over 5})$, in the flat space ($k = 0$). Thus, Einstein's scalar-tensor equivalent frame corresponding to $F(R)$ theory of gravity appears to be more general, since it leads to several conserved currents, which were absent otherwise. On one hand, while it was encouraging to recover the form $F(R) \propto R^2$, unfortunately, on the other, the form $F(R) \propto R^{\frac{3}{2}}$ remains obscure in Einstein's frame. It therefore appeared that not only the frames are inequivalent at-least in the context of Noether symmetry, but also symmetries remain hidden in different frames \cite{6}. Nevertheless, it was soon realized that the root cause for this apparent in-equivalence and hidden symmetries, might be due to the fact that Noether theorem is not on-shell for constrained system such as gravity, being associated with diffeomorphic invariance \cite{7}. For this reason, it is always suggestive to systematically check, if the symmetry obtained in view of Noether symmetry analysis satisfies the field equations, particularly the ($^0_0$) equations of Einstein, in the absence of space-time components of the metric \cite{We1, We2, We3, We4}. Otherwise, if one wants to look for on-shell symmetry, avoiding the task of checking the field equations, and wants to explore all hidden symmtries, it is mandatory to involve the constraints in the symmetry generator \cite{7}. Consequently, we proposed a symmetry generator in the form,

\be\label{gen} \pounds_X L -\eta E = X L - \eta E = 0,\ee
where $X$ is the vector field, $L$ is the point Lagrangian, and $E = 0$ represents the ($^0_0$) equation (the energy equation) of Einstein, which is essentially the Hamiltonian, when expressed in terms of phase-space variables, and is constrained to vanish due to diffeomorphic invariance. In the above relation \eqref{gen}, $\eta$ acts as a Lagrange multiplier, being a function of generalised coordinates \footnote{It must be mentioned that three momenta constraints are involved for a general metric due to the appearance of space-time components of the metric. Thus \eqref{gen} should contain $\delta_i P^i$ in addition, $P^i$ being the momenta and $\delta_i$ are Lagrange multipliers}. For certain choices of $\eta$ it had been possible to find all the four symmetries in all the frames, establishing equivalence as well \cite{7}. \\

The chapter is not closed as yet, since as already stated, modified (by $F(R)$) scalar-tensor theory admits arbitrary form of $F(R)$, which is explored from the field equations, as mentioned, but could not be associated with Noether symmetry, as yet. On the other hand, for pure $F(R)$ theory, only a few choices of the Lagrange multiplier $\eta$ have been administered so far, since the aim was just to establish equivalence between different frames \cite{7}. However, other choices might also lead to even more symmetries. In fact, some authors claimed to have obtained a host of symmetries following different techniques. For example, it was claimed by Hussain et-al \cite{8} that Noether gauge symmetry exclusively for $F(R)$ theory admits $F(R) \propto R^n$, where $n$ is arbitrary, while Jamil et-al \cite{9} found $F(R) \propto R^2$ and the potential, $V(\phi) \propto \phi^{-4}$, considering Noether gauge symmetry with Born-Infeld-Tachyonic action. Both the claims had been reviewed by the present authors and found to be incorrect, since they don't satisfy the field equation \cite{10, 11, 4c}. Date back, Shamir et-al \cite{12} claimed that in the presence of non-zero gauge (although it's essentially a boundary term, and not a gauge) Noether symmetry of $F(R) \propto R^{\frac{3}{2}}$ admits four different generators corresponding to which four different conserved currents exists. In the mean-time, Paliathanasis et al \cite{13} and later Paliathanasis \cite{14} had further obtained a host of different forms of $F(R)$ and their corresponding conserved currents, in view of Killing tensors, some of which may have interesting cosmological consequences.  The aim of the present manuscript is to check the viability of these claims \cite{12, 13, 14} from a different perspective, viz. in view of the first Poisson theorem.\\

Poisson theorem of classical mechanics states that: The total time derivative of a function of phase-space variables is the sum of its poisson bracket with the Hamiltonian and its partial time derivative. Thus, if the function does not contain time explicitly, then its total time derivative is equal to its Poisson bracket with the Hamiltonian, and it is conserved if the poisson bracket vanishes. With the help of Poisson theorem, it is possible to perform a general test for seeking and identifying the constants of motion. In the following section, we express the point Lagrangian associated with $F(R)$ theory of gravity taking the scale factor and the Ricci scalar along with the generalized velocities ($a, R, \dot a, \dot R$) in the background of Robertson-Walker line element, write down the energy equation, viz. the ($^0_0$) equation of Einstein, cast the phase-space structure of the Hamiltonian, and provide a brief account of the well-known Poisson theorem. In section 3, we proceed to test the conserved currents of the theory available in the literature, in view of the said Poisson theorem. In section 4, we provide a list of available (viable) forms of $F(R)$ with admissible symmetries, presently available in the literature. In view of such forms, we construct a generalised action, which might possibly be able to address the cosmic puzzle, elegantly. Finally, we conclude in section 5.

\section{Lagrangian and Hamiltonian formulation of F(R) gravity.}

Taking Robertson-Walker line element,

\be ds^2 = -dt^2+a^2\left[\frac{dr^2}{1-kr^2} + r^2 \big(d\theta^2 + r^2\sin^2\theta d\phi^2\big)\right],\ee

\noindent
into account, the following modified theory of gravitational action for an arbitrary function of $R$, viz.

\be A = \int d^4 x \sqrt{-g}\left[ F(R) + \mathcal{L}_m \right],\ee
may be expressed as

\be A = \int d^4 x \sqrt{-g}\left[ F(R) + \lambda\left\{R - 6\left(\frac{\ddot a}{a} + \frac{\dot a^2}{a^2} + \frac{k}{a^2}\right)\right\} \right] + \int d^4 x \sqrt{-g}\mathcal{L}_m,\ee
where, the Ricci scalar curvature, $R - 6\left(\frac{\ddot a}{a} + \frac{\dot a^2}{a^2} + \frac{k}{a^2}\right) = 0$
has been treated as a constraint of the theory, and inserted through the Lagrange multiplier $\lambda$, while the Lagrangian is spanned by a set of configuration space variable $(a, R, \dot a, \dot R)$. Now varying the above action with respect to $R$, one obtains $\lambda = F'$, where prime denotes derivative with respect to the Ricci scalar $R$. The above action therefore reads as,

\be A = \int d^4 x \sqrt{-g}\left[F(R) a^3 + F'R a^3  +  6F'\left(a^2\ddot a + a\dot a^2 + ka\right) \right] + \int d^4 x \sqrt{-g}\mathcal{L}_m.\ee
Under integration by parts, one therefore ends up with the point Lagrangian,

\be\label{L} L(a, R, \dot a, \dot R)= -6a \dot a^2 F'  - 6a^2\dot a\dot R F'' - a^3(F' R - F) + 6kaF' - \rho_0 a^{-3\omega}.\ee

\noindent
In the above relations, $\mathcal{L}_m = \rho_0 a^{-3(\omega + 1)}$ denotes matter Lagrangian density, $\rho_0$ being the present value of the matter density either in the form of radiation ($\rho_{r0}$), or pressure-less dust ($\rho_{m0}$), while it vanishes ($\rho_0 = 0$) in vacuum, and $\omega$ is the state parameter ($1/3$ for radiation, $0$ for pressure-less dust). The field equations together with the following energy constraint equation, viz. the ($^0_0$) equation of Einstein, are the following,

\be \label{E} E = -6a \dot a^2 F'  - 6a^2\dot a\dot R F'' + a^3(F' R - F) - 6 k a F' + \rho_0 a^{-3\omega} = 0,\ee
are expressed as,

\be\begin{split} \label{FE1} &2{\ddot a\over a} + {\dot a^2\over a^2} + {k\over a^2} + {F''\over F'}\left(\ddot R + 2{\dot a\over a}\dot R\right) + {F'''\over F'}\dot R^2 - {1\over 2F'} (F'R - F) = - {\omega\rho_0\over 2 F'} a^{-3(\omega+1)},\\&
R = 6\left(\frac{\ddot a}{a} + \frac{\dot a^2}{a^2} + \frac{k}{a^2}\right),\\&
\frac{\dot a^2}{a^2} + \frac{k}{a^2} + {F''\over F'}\left({\dot a \over a}\right)\dot R - {1\over 6F'} (F'R - F) = {\rho_0\over 6 F'} a^{-{3(\omega+1)}}.\end{split}\ee
For our future requirement, we also write the expression for the point Lagrangian in Jordan's frame as \cite{6, 7},

\be\label{L1} L(a, \phi, \dot a, \dot {\phi}) =- 6a \dot a^2 \phi  - 6a^2\dot a\dot {\phi} - a^3 V(\phi)+ 6ka\phi - \rho_0 a^{-3\omega},\ee
where, $\phi = F'(R)$, and $V(\phi) = F'(R)R - F(R)$ is the first-order Clairaut equation. Here again, the field equations together with the following energy constraint equation,

\be\label{E1} E = -6a \dot a^2 \phi  - 6a^2\dot a\dot\phi + a^3(V(\phi) - 6 k a \phi + \rho_0 a^{-3\omega} = 0,\ee
are expressed as,

\be\begin{split} \label{FE2} &2{\ddot a\over a} + {\dot a^2\over a^2} + {k\over a^2} + {\ddot \phi\over \phi} + 2{\dot a\over a}{\dot \phi\over \phi} - {V\over 2\phi}  = - {\omega\rho_0\over 2 \phi} a^{-3(\omega+1)},\\&
V_{,\phi} = 6\left(\frac{\ddot a}{a} + \frac{\dot a^2}{a^2} + \frac{k}{a^2}\right),\\&
\frac{\dot a^2}{a^2} + \frac{k}{a^2} + {\dot a \dot\phi\over a\phi}- {V\over 6\phi} = {\rho_0\over 6 \phi} a^{-{3(\omega+1)}}.\end{split}\ee\\

\noindent
\textbf{The Hamiltonian:}\\

\noindent
The generic momenta with respect to variables $a$ and $R$ are,

\be\label{CM1} P_{a} = \frac{\partial L}{\partial \dot{a}}= - 12a\dot{a}F' - 6 a^2\dot{R}F'';~~~P_{R} =  \frac{\partial L}{\partial \dot{R}}= -6 a^2\dot{a}F''.\ee
Thus we have,
\be \label{GC1}\dot{a} = -\frac{P_{R}}{6a^2F''}; ~~~\dot{R} = \frac{2F'P_{R} - a F''P_{a}}{6a^3{F''}^2}.\ee
The Hamiltonian corresponding to the point Lagrangian \eqref{L} can then be expressed as,
\be \label{H1} H(a, R, P_{a}, P_{R})= P_{a}\dot{a} + P_{R}\dot{R}-L = \frac{F'P_{R}^2}{6a^3{F''}^2} - \frac{P_{a}P_{R}}{6a^2F''} + a^3(F' R - F) -6k a F' + \rho_0 a^{-3\omega} = 0,\ee
which is constrained to vanish due to diffeomorphic invariance. Likewise, the Hamiltonian in Jordan's frame reads as,

\be \label{H2} H(a, \phi, P_{a}, P_{\phi}) = \frac{\phi P_{\phi}^2}{6a^3} - \frac{P_{a}P_{\phi}}{6a^2} + a^3 V(\phi) -6k a \phi + \rho_0 a^{-3\omega} = 0.\ee

\noindent
\textbf{Poisson Theorem:} \\

\noindent
Conserved currents play an important role in theoretical physics, since the existence of a conserved current points to the existence of a constant of motion and a symmetry. The Poisson theorem ascertains if indeed a function is a constant of motion of the theory under consideration. Let, $I = I(q_i, p_i, t)$ be an arbitrary function of phase space variables. So,

\be \label{PT1}\frac {dI}{dt} = \sum_{i}\left(\frac{\partial I}{\partial q_{i}}\dot{q_i}+ \frac{\partial I}{\partial p_{i}}\dot{p_i}\right)+\frac{\partial I}{\partial {t}}.\ee
Using Hamilton's equations, in the above relation, one obtains the Poisson theorem in the following form,

\be \label{P}\frac {dI}{dt} = \sum_{i}\left(\frac{\partial I}{\partial q_{i}}\frac{\partial H}{\partial p_{i}}- \frac{\partial I}{\partial p_{i}}\frac{\partial {H}}{\partial q_{i}}\right)+\frac{\partial I}{\partial {t}}=[I, H]+\frac{\partial I}{\partial {t}}.\ee
Thus, if the function $I$ does not contain time explicitly (${\partial I\over \partial t} = 0$), then its total time derivative is equal to its poisson bracket with the Hamiltonian (${dI\over dt} = [I, H]$), and it is conserved if the poisson bracket vanishes. The theorem of classical mechanics should be modified in the case of gravity, where the Hamiltonian is constrained to vanish due to diffemorphic invariance. The theorem now states \emph{`If the right hand side vanishes identically or is proportional to the Hamiltonian, then $I$ is an integral of motion'}. On the contrary, we shall show in the appendix that if the right hand side vanishes conditionally, the field equations are not satisfied and $I$ is not an integral of motion.\\

\noindent
As mentioned, for constrained system such as gravity, Noether symmetry is not on-shell, and setting the Lie derivative of the Lagrangian with respect to a vector field to vanish, might exhibit symmetries, which are not the symmetries of the system in one hand, and some of the symmetries might remain hidden, on the other. On the contrary, although, all the symmetries exhibited following the generator \eqref{gen} are the symmetries of the system, it is a rather difficult technique, since one is required to set $\eta$ by hand using trial and error method to explore symmetry. However, following some elegant technique other than considering the generator \eqref{gen}, if new symmetries appear, then one has to check if the symmetries satisfy the field equation, the constraint equation in particular, since as mentioned, Noether equations do not recognise the constraints of a theory. In this context, it is easiest to check if indeed these are the symmetries of the system, with the help of Poisson theorem. Thus Poisson theorem renders a general test for seeking and identifying the constants of motion. In the following sections, we apply Poisson theorem to check the viability of all the conserved currents associated with $F(R)$ theory of gravity, available in the literature to the best of our knowledge, and in the process, we rule out quite a large number of symmetries .

\section{Testing Noether Conserved currents in F(R) theory of gravity:}

As already mentioned, Noether's theorem connects the existence of a conserved current to the existence of a symmetry of the system under consideration. In this section, we provide explicit test of the Noether conserved currents associated with $F(R)$ theory of gravity, which are available in the literature till date. In the present article, the phase-space is spanned in most of the cases by ($a, R, P_a, P_R$), whence the Poisson theorem reads as,

\be\label{P1} \frac{dI}{dt} = [I, H] +\frac{\partial I}{\partial {t}} = \left(\frac{\partial {I}}{\partial a}\frac{\partial H}{\partial P_{a}}- \frac{\partial {I}}{\partial P_{a}}\frac{\partial {H}}{\partial a}\right) + \left(\frac{\partial {I}}{\partial R}\frac{\partial H}{\partial P_{R}}- \frac{\partial{I}}{\partial P_{R}}\frac{\partial {H}}{\partial R}\right) +\frac{\partial I}{\partial {t}}.\ee
In one case, to avoid unnecessary complication, we work in Jordan's frame of reference, in which the phase-space is spanned by ($a,\phi,P_a,P_\phi$), whence the Poisson theorem reads as,

\be \label{P2} \frac{d I}{dt} = [I, H] +\frac{\partial I}{\partial {t}}= \left(\frac{\partial {I}}{\partial a}\frac{\partial H}{\partial P_{a}}- \frac{\partial {I}}{\partial P_{a}}\frac{\partial {H}}{\partial a}\right) + \left(\frac{\partial {I}}{\partial \phi}\frac{\partial H}{\partial P_{\phi}}- \frac{\partial{I}}{\partial P_{\phi}}\frac{\partial {H}}{\partial \phi}\right) +\frac{\partial I}{\partial {t}}.\ee

\subsection{Conserved currents provided in \cite{7}:}

In an earlier work, \cite{7}, altogether $4$ different forms of $F(R)$ along with the corresponding conserved currents have been found upon Noether symmetry analysis of $F(R)$ theory of gravity, using Lagrange multiplier technique, translating it to the scalar-tensor equivalent form in Jordan's and in Einstein's frames. This fact established the equivalence amongst different frames in connection with Noether symmetry. All the available forms along with their corresponding conserved currents were found to satisfy the field equations, particularly the $(^0_0)$ equation of Einstein. Still, to see how Poisson theorem works, we make a rapid test of these conserved currents in this subsection.\\

\noindent
\textbf{Case-I:}\\

\noindent
The first form of $F(R)$ and the associated conserved charge were found both in the vacuum era ($p = 0 = \rho$) and matter dominated era ($p = 0$, or equivalently, $\omega = 0$) for $k = 0, \pm 1$ as,
\be \label{sigma} F(R) = F_0 R^{\frac{3}{2}},\hspace{1.0 cm}{\Sigma_{1}} =\frac{d}{dt} ({a\sqrt R}) = \dot{a}\sqrt{R}+\frac{a\dot{R}}{2\sqrt{R}}.\ee
In view of the above form of $F(R)$, the Hamiltonian \eqref{H1}, and the phase-space structure of the above conserved current ${\Sigma_{1}}$ using equations \eqref{GC1}, may be expressed as,

\be\label{H-1} \begin{split}&H_{1}=\frac{4R^\frac{3}{2} P_{R}^2}{9 F_0 a^3} - \frac{2R^\frac{1}{2}P_{a}P_{R}}{9 F_0 a^2} + \frac{1}{2}F_0 a^3R^\frac{3}{2} - 9kF_0 a \sqrt R + \rho_0,\hspace{0.80 cm}{\Sigma_{1}} =\frac{1}{F_0}\left(\frac{2RP_{R}}{9a^2}-\frac{P_{a}}{9a}\right),\end{split}\ee
where we have kept the term $\rho_0$, to accommodate both the vacuum ($\rho_0 = 0$) as well as the matter $\rho_0 = \rho_{m0}$ for pressure-less dust ($\omega = 0$). As a result one can compute,

\be\begin{split}\label{HD1}&
\frac{\partial {H_{1}}}{\partial {a}} = \frac{4R^\frac{1}{2}P_a P_R}{9 F_0 a^3} - \frac{4R^\frac{3}{2} P_R^2}{3F_0 a^4} + \frac{3F_0 a^2R^\frac{3}{2}}{2} - 9k F_0 \sqrt R,\hspace{1.45 cm}
\frac{\partial {H_{1}}}{\partial {P_a}} = - \frac{2R^\frac{1}{2} P_R}{9F_0 a^2}.\\&
\frac{\partial {H_{1}}}{\partial {R}} = \frac{2R^\frac{1}{2} P_R^2}{3 F_0 a^3} - \frac{R^{-\frac{1}{2}}P_a P_R}{9F_0 a^2} + \frac{3 F_0 a^3R^\frac{1}{2}}{4} - {9 k a F_0\over 2 \sqrt R},\hspace{1.7 cm}\frac{\partial {H_{1}}}{\partial {P_R}} = -\frac{2R^\frac{1}{2} P_a}{9 F_0 a^2} + \frac{8R^\frac{3}{2} P_R}{9 F_0a^3}.\\&
\frac{\partial {\Sigma_{1}}}{\partial {a}} = \frac{P_a}{9F_0 a^2}-\frac{4RP_R}{9F_0 a^3},\hspace{5.98 cm}\frac{\partial {\Sigma_{1}}}{\partial {P_a}} = -\frac{1}{9F_0 a}.\\&
\frac{\partial {\Sigma_{1}}}{\partial {R}} = \frac{2P_R}{9F_0 a^2},\hspace{7.38 cm}\frac{\partial {\Sigma_{1}}}{\partial {P_R}} = \frac{2R}{9F_0 a^2}.
\end{split}\ee
Above relations hold both in the vacuum $(\rho_0 = 0)$ and in pressure-less dust era ($\rho_{m0}$), and it is now possible to check in view of \eqref{P1} that,

\be \frac {d{\Sigma_{1}}}{dt} = 0,\ee
which confirms that $\Sigma_1$ is indeed the conserved current associated with the Hamiltonian $H_1$, and hence Noether symmetry allows the form $F(R) \propto R^{3\over 2}$. It is important to mention that since the Poisson bracket vanishes trivially, so it corresponds to the case $\eta = 0$ in equation \eqref{gen}, as found in \cite{7}.\\

\noindent
\textbf{Case-II:}\\

\noindent
The next form of $F(R)$ and the associated conserved current were obtained both in vacuum and in the radiation dominated era for $k = 0, \pm 1$ as,
\be \label{R2} F(R) = F_0 R^2,\hspace{1.0 cm}{\Sigma_{2}}  =  a^3\dot R. \ee
In view of the above form of $F(R)$, the Hamiltonian \eqref{H1}, and the phase-space structure of the above conserved current ${\Sigma_{2}}$ using equations \eqref{GC1}, may be expressed as,

\be H_{2}= \frac{R P_{R}^2}{12F_0 a^3} -\frac{P_{a}P_{R}}{12 F_0 a^2} + F_0 a^3R^2 - 12 k F_0 a R + \rho_0 a^{-3\omega},\hspace{1.3 cm}{\Sigma_{2}}  = \frac{1}{12 F_0}(2R P_{R}- a P_{a}).\ee
Here we keep the matter term $\rho_0 a^{-3\omega}$ to accommodate all the different eras together. One can thus compute,

\be\begin{split}&
\frac{\partial {H_{2}}}{\partial {a}} = \frac{R P_R^2}{4F_0 a^4}+\frac{P_a P_R}{6F_0 a^3}+3F_0 a^2R^2 - 12kF_0R - 3\rho_0\omega a^{-(3\omega -1)},\hspace{0.2 cm}\frac{\partial {H_{2}}}{\partial {P_a}}= -\frac{P_R}{12F_0 a^2},\\&
\frac{\partial {H_{2}}}{\partial {R}} = \frac{P_R^2}{12F_0 a^3} + 2F_0 a^3R - 12kF_0 a,\hspace{4.18 cm}\frac{\partial {H_{2}}}{\partial {P_R}}= \frac{R P_R}{6F_0 a^3} -\frac{P_a}{12F_0 a^2} ,\\&
\frac{\partial {\Sigma_{2}}}{\partial {a}} = -\frac{P_a}{12F_0},\hspace{7.44 cm}\frac{\partial {\Sigma_{2}}}{\partial {P_a}}= - \frac{a}{12F_0},\\&
\frac{\partial {\Sigma_{2}}}{\partial {R}} = \frac{P_R}{6F_0},\hspace{7.9 cm}\frac{\partial {\Sigma_{2}}}{\partial {P_R}}= \frac{R}{6F_0},
\end{split}\ee
and in view of the above relations, one can at once check in view of \eqref{P1} that,

\be \begin{split}\frac {d{\Sigma_{2}}}{dt} = \frac{1}{12 F_0}\left(\frac{R P_{R}^2}{12F_0 a^3} - \frac{P_{a}P_{R}}{12F_0 a^2} + F_0 a^3R^2 -12kF_0 a R + 3\rho_0\omega a ^{-3\omega}\right).
\end{split}\ee
It is to be noted fact that the above Poisson bracket reduces to $[{\Sigma_{2}},H_{2}]=\frac{H_{2}}{12 F_0}$ both for $\rho_0 = 0$, i.e. in vacuum, and for $\omega = {1\over 3}$, i.e. for radiation era, while it is $[{\Sigma_{2}},H_{2}]=\frac{H_{2}}{12 F_0} - \rho_0$ in the matter dominated era. Thus the Poisson bracket vanishes due to the fact that the Hamiltonian is constrained to vanish due to diffeomorphic invariance both in the vacuum and in the radiation dominated eras, but not in the matter dominated era. This confirms that $\Sigma_2$ is indeed the conserved current associated with the Hamiltonian $H_2$, and Noether symmetry allows $F(R) \propto R^2$ both in the vacuum and the radiation dominated eras. It is important to mention that the Poisson bracket vanishes due to the diffeomorphic invariance, which is associated with $\eta = 1$, requiring ${12 F_0} = 1$, to match exactly with \cite{7}.\\

\noindent
\textbf{Case-III:}\\

\noindent
The third form of $F(R)$ and the associated conserved charge were obtained in vacuum era alone and in the flat space ($k = 0$) as,

\be \label{-1} F(R) =\frac{ F_0}{R},\hspace{1.0 cm} {\Sigma_{3}} = R\dot a\sqrt a.\ee
Associated Hamiltonian in view of equation \eqref{H1} and the conserved current ${\Sigma_{3}}$ using equation \eqref{GC1} are expressed as,

\be H_{3}= -\frac{R^3 P_{a}P_{R}}{12F_0 a^2}-\frac{R^4 P_{R}^2}{24 F_0 a^3}-\frac{2F_0 a^3}{R}, \hspace{1.2 in}{\Sigma_{3}} = -\frac{R^4 P_{R}}{12 F_0 a^\frac{3}{2}},\ee
whose derivatives read as,

\be\begin{split}&
\frac{\partial {H_{3}}}{\partial {a}} = \frac{R^3 P_{a}P_{R}}{6F_0 a^3}+\frac{R^4 P_{R}^2}{8F_0 a^4}-\frac{6F_0 a^2}{R},\hspace{1.22 in}\frac{\partial {H_{3}}}{\partial {P_a}}= -\frac{R^3 P_{R}}{12 F_0a^2}.\\&
\frac{\partial {H_{3}}}{\partial {R}} = -\frac{R^2 P_{a}P_{R}}{4F_0 a^2} - \frac{R^3 P_{R}^2}{6 F_0a^3} + \frac{2F_0 a^3}{R^2},\hspace{1.11 in}
\frac{\partial {H_{3}}}{\partial {P_R}} = -\frac{R^3 P_{a}}{12F_0 a^2} - \frac{R^4 P_{R}}{12F_0 a^3}.\\&
\frac{\partial {\Sigma_{3}}}{\partial {a}} = \frac{R^4 P_{R}}{8 F_0a^\frac{5}{2}},\hspace{2.478 in}\frac{\partial {\Sigma_{3}}}{\partial {P_a}}= 0.\\&
\frac{\partial {\Sigma_{3}}}{\partial {R}} = -\frac{R^3 P_{R}}{3F_0 a^\frac{3}{2}},\hspace{2.37 in}\frac{\partial {\Sigma_{3}}}{\partial {P_R}}=-\frac{R^4}{12 F_0a^\frac{3}{2}}.
\end{split}\ee
In view of the above relations, one can at once check that,

\be\begin{split} \frac {d{\Sigma_{3}}}{dt} = \frac{R^3}{12F_0a^\frac{3}{2}}\left(- \frac{R^3 P_{a}P_{R}}{12F_0 a^2} - \frac{R^4 P_{R}^2}{24 F_0a^3} - \frac{2F_0 a^3}{R}\right) = \frac{R^3 H_{3}}{12F_0 a^\frac{3}{2}}= 0.\end{split}\ee
Here again the Poisson bracket vanishes due to diffeomorphic invariance, and thus confirming that $\Sigma_3$ is also the conserved current associated with the Hamiltonian $H_3$, and Noether symmetry allows $F(R) \propto R^{-1}$ in vacuum. Note that this conserved current was obtained earlier for $\eta \propto {R^3 a^{-{3\over 2}}}$ \cite{7}, which is the coefficient of the Hamiltonian here, as seen in the last step. To match exactly, one should choose $\beta_0 = -1$, in \cite{7} and $12 F_0 = 1$ here.\\

\noindent
\textbf{Case-IV:}\\

\noindent
The fourth and the final form of $F(R)$ and the associated conserved charge were also obtained in vacuum dominated era for $k = 0$ as,

\be \label{75} F(R) = F_{0}R^{\frac{7}{5}},\hspace{0.5 in} {\Sigma_{4}} = \sqrt a {d\over dt}(a R^{2\over 5}) = R^{-{3\over 5}}(R\sqrt a \dot a + {2\over 5} a^{3\over 2} \dot R).\ee
The Hamiltonian as well as the phase-space structure of the above conserved current can then be expressed using equation \eqref{H1} and \eqref{GC1} respectively as,

\be H_{4}= \frac{125R^\frac{8}{5} P_{R}^2}{168 F_0a^3} - \frac{25 R^\frac{3}{5} P_{a}P_{R}}{84F_0 a^2} + \frac{2F_0 a^3 R\frac{7}{5}}{5},\hspace{0.62 in}{\Sigma_{4}}= {1\over F_0}\left(\frac{25RP_{R}}{84 a^{3\over 2}} -\frac{5P_{a}}{42\sqrt{a}}\right),\ee
while their derivatives are computed as,

\be\begin{split}&
\frac{\partial {H_{4}}}{\partial {a}} = -\frac{125 R^\frac{8}{5} P_{R}^2}{56F_0a^4} +\frac{25R^\frac{3}{5} P_{a}P_{R}}{42F_0 a^3} + \frac{6F_0a^2R^\frac{7}{5}}{5},\hspace{0.31 in}\frac{\partial {H_{4}}}{\partial {P_a}} = -\frac{25R^\frac{3}{5}P_{R}}{84 F_0a^2}.\\&
\frac{\partial {H_{4}}}{\partial {R}} =\frac{25 R^\frac{3}{5} P_{R}^2}{21F_0 a^3} - \frac{5R^{-\frac{2}{5}} P_{a}P_{R}}{28 F_0a^2} + \frac{14F_0a^3R^\frac{2}{5}}{25},\hspace{0.4 in}\frac{\partial {H_{4}}}{\partial {P_R}}= \frac{125R^\frac{8}{5}P_{R}}{84 F_0 a^3} - \frac{25R^\frac{3}{5}P_{a}}{84F_0 a^2}.\\&
\frac{\partial {\Sigma_{4}}}{\partial {a}} = \frac{5 P_a}{84 F_0a^\frac{3}{2}} - \frac{25 R P_R}{56 F_0 a^\frac{5}{2}},\hspace{1.54 in}\frac{\partial {\Sigma_{4}}}{\partial {P_a}}= -\frac{5}{42 F_0a^\frac{1}{2}}.\\&
\frac{\partial {\Sigma_{4}}}{\partial {R}} =  \frac{25P_R}{84 F_0a^\frac{3}{2}},\hspace{2.18 in}\frac{\partial {\Sigma_{4}}}{\partial {P_R}}=  \frac{25 R}{84 F_0a^\frac{3}{2}}.
\end{split}\ee
In view of the above relations, it is now possible to find as before,

\be \begin{split}\frac {d{\Sigma_{4}}}{dt} = \frac{5}{84F_0 a^\frac{3}{2}}\left(\frac{125R^\frac{8}{5} P_{R}^2}{168F_0 a^3}- \frac{25 R^\frac{3}{5} P_{a}P_{R}}{84F_0 a^2} + \frac{2F_0 a^3 R\frac{7}{5}}{5}\right)=\frac{5 H_{4}}{84F_0a^\frac{3}{2}}= 0.\end{split}\ee
Diffeomorphic invariance is the reason for the vanishing of the above Poisson bracket as observed in the two previous cases. Thus, $\Sigma_4$ is also the conserved current associated with the Hamiltonian $H_4$, while Noether symmetry allows the form $F(R) \propto R^{7\over 5}$ in vacuum. Note that this conserved current was obtained for $\eta \propto {a^{-{3\over 2}}}$, which is the coefficient of the Hamiltonian as seen in the last step. In fact it matches exactly under the choice $\alpha_0 = {5\over 7}$ in \cite{7} and $12 F_0 = 1$, as before. As mentioned, inclusion of the energy constraint in the generator, makes Noether symmetry on-shell. Further, since right hand side of the Poisson theorem \eqref{P} either vanishes or is proportional to the Hamiltonian, hence field equations are satisfied automatically.

\subsection{Conserved currents provided in \cite{12}:}

Almost a decade back, applying Noether symmetry generator in the following form

\be \begin{split} &X^{[1]}L + L {d\tau\over dt} - {d B\over dt} = 0,~\mathrm{where},\\&
X^{[1]} = X + \dot \psi(t, a, R){\partial\over \partial \dot a} + \dot \phi(t, a, R){\partial\over \partial \dot R},~\mathrm{is ~the~first~prolongation~ of,}\\&
X = \tau(t, a, R){\partial\over \partial t} + \psi(t, a, R){\partial\over \partial a} + \phi(t, a, R){\partial\over \partial a},\end{split}\ee
Shamir et-al \cite{12} explored following four different conserved currents for the theory under consideration, corresponding to $F(R) \propto R^{\frac{3}{2}}$ in flat space ($k = 0$) and in the vacuum dominated era,

\be \label{SI} \begin{split} &I_1 = -3a^2\dot a \sqrt R- 3a^3\frac{\dot R}{\sqrt R}+\frac{t}{2}\left[18 a\dot a^2 \sqrt R+9 a^2\dot a\frac{\dot R}{\sqrt R}+a^3 R^\frac{3}{2}\right],\\&
I_2 = -9a\dot a^2 \sqrt R-\frac{9}{2}a^2\dot a\frac{\dot R}{\sqrt R}+\frac{1}{2}a^3 tR^{\frac{3}{2}},\\&
I_3 = 9a\sqrt R-t\Big[9\dot a\sqrt R+\frac{9a}{2}\frac{\dot R}{R^{\frac{1}{2}}}\Big]=9\left[{a\sqrt R}-t \frac{d}{dt} ({a\sqrt R})\right],\\&
I_4 = -9\dot a\sqrt R-\frac{9 a}{2}\frac{\dot R}{\sqrt R}=-9\frac{d}{dt} ({a\sqrt R}). \end{split}\ee
In the above, $B$ although claimed to be, is not a gauge function, rather a boundary term as pointed out earlier \cite{12}. It may be noted that $I_4$ is the same conserved current ($\Sigma_1$) dealt with in case-1 \eqref{sigma} of subsection (3.1). Further since,

\be {d I_3\over dt} = -9t\left[{d^2(a\sqrt R)\over dt^2}\right]  = - 9t \left[{d\Sigma_1\over dt}\right],\ee
it is also not a new conserved current. Thus in this subsection we only handle the first two conserved currents, whose phase-space structures may be expressed using \eqref{GC1} as,

\be \label{SI2} \begin{split} &{I_{1}}=\left[-\frac{4R^\frac{3}{2} P_{R}^2}{9 F^2_0 a^3} + \frac{2R^\frac{1}{2}P_{a}P_{R}}{9 F^2_0 a^2} + \frac{1}{2} a^3R^\frac{3}{2}\right] t+ \frac{2a P_{a}}{3F_{0}}-\frac{2R P_{R}}{F_{0}},\\&
{I_{2}}=\frac{4R^\frac{3}{2} P_R^2 }{9F^2_0a^3}-\frac{2R^{\frac{1}{2}} P_{a} P_{R}}{9F^2_{0}a^2}+\frac{a^3R^{\frac{3}{2}}}{2} t.\end{split}\ee
The corresponding Hamiltonian and its derivatives are given in \eqref{H-1} and \eqref{HD1} respectively.\\

\noindent
\textbf{Case-I:}\\

\noindent
We take the first conserved current appearing in \eqref{SI2} and find its derivatives as,
\be\begin{split}&
\frac{\partial {I_{1}}}{\partial {a}}= t\left[-\frac{4R^\frac{1}{2}P_a P_R}{9 F^2_0 a^3} + \frac{4R^\frac{3}{2} P_R^2}{3F^2_0 a^4} + \frac{3 a^2R^\frac{3}{2}}{2}\right]+\frac{2 P_{a}}{3F_{0}},
\hspace{0.26 in}\frac{\partial {I_{1}}}{\partial {P_a}}=t\left[ \frac{2R^\frac{1}{2} P_R}{9F^2_0 a^2}\right]+ \frac{2a}{3 F_{0}},\\&
\frac{\partial {I_{1}}}{\partial {R}}=t\left[-\frac{2R^\frac{1}{2} P_R^2}{3 F^2_0 a^3} + \frac{R^{-\frac{1}{2}}P_a P_R}{9F^2_0 a^2} + \frac{3 a^3R^\frac{1}{2}}{4}\right] -\frac{2P_{R}}{ F_{0}},
\hspace{0.2 in}
\frac{\partial {I_{1}}}{\partial {P_R}}=t\left[ \frac{2R^\frac{1}{2} P_a}{9 F^2_0 a^2} - \frac{8R^\frac{3}{2} P_R}{9 F_0^2 a^3}\right] -\frac{2R}{F_{0}},\\&
\frac{\partial {I_{1}}}{\partial t} = -\frac{4R^\frac{3}{2} P_{R}^2}{9 F^2_0 a^3} + \frac{2R^\frac{1}{2}P_{a}P_{R}}{9 F_0^2 a^2} + \frac{1}{2} a^3R^\frac{3}{2}.
\end{split}\ee
One can therefore compute the time derivative of the conserved current using the Poisson theorem \eqref{P1} as,
\be \frac {d{I_{1}}}{dt} = [{I_{1}},H]+\frac{\partial {I_{1}}}{\partial {t}} = t\left[\frac{2R^2}{3 F_0}P_R-\frac{ a R  }{3F_0 }P_a\right]
+ a^3R^\frac{3}{2}\ne0.\ee
Thus, it is a faulty claim that, $I_1$ is a conserved current associated with the $F(R) \propto R^{3\over2}$ in \cite{12}. \textbf{It is not}.\\

\noindent
\textbf{Case-II:}\\

\noindent
Derivatives of the second conserved current appearing in \eqref{SI2} are,

\be\begin{split}&
\frac{\partial {I_{2}}}{\partial {a}}= \frac{3a^2R^{\frac{3}{2}}}{2} t+\frac{4R^{\frac{1}{2}} P_{a} P_{R}}{9a^3 F^2_{0}}-\frac{4R^\frac{3}{2} P^2_R}{3F^2_0 a^4}, \hspace{0.34 in} \frac{\partial {I_{2}}}{\partial {P_a}}= -\frac{2R^{\frac{1}{2}} P_{R}}{9a^2 F^2_{0}},\\&
\frac{\partial {I_{2}}}{\partial {R}}= \frac{3a^3R^{\frac{1}{2}}}{4}t - \frac{R^{-\frac{1}{2}} P_{a} P_{R}}{9F^2_{0}a^2 } +\frac{2R^\frac{1}{2}P^2_R}{3F^2_0 a^3},\hspace{0.3 in}
\frac{\partial {I_{2}}}{\partial {P_R}}= \frac{8R^\frac{3}{2}P_R}{9F^2_0 a^3}-\frac{2R^{\frac{1}{2}} P_{a}}{9a^2 F^2_{0}},\hspace{0.2 in}
\frac{\partial {I_{2}}}{\partial {t}}= \frac{a^3R^{\frac{3}{2}}}{2}.
\end{split}\ee
Thus, one ends up with,

\be\begin{split} &\frac {d{I_{2}}}{dt} = [{I_{2}},H]+\frac{\partial {I_{2}}}{\partial {t}} = (t-1)\left[\frac{R^2 P_{R}}{3F_0}-\frac{a R P_{a}}{6F_0}\right]+\frac{a^3 R^{\frac{3}{2}}}{2} \neq 0.\end{split}\ee
Clearly, \textbf{the claim that $I_2$ is also a conserved current associated with $F(R) \propto R^{3\over 2}$ is faulty}. \\

\noindent
As mentioned, Noether equations do not recognise the existing constraints of a theory. Root of trouble is: as mentioned, in the case of gravity, if the energy constraint is not introduced in the symmetry generator, one is supposed to check the field equations, particularly the constraint equation, which was not performed by the authors \cite{12}. This was also noticed earlier, where it was shown that the conserved current $I_2$ doesn't satisfy the field equations, viz. the ($^0_0$) equation of Einstein, in particular \cite{11}. It is therefore quite clear that introduction of the so-called gauge term ($B$) does not yield any new symmetry, and thus the work of Shamir et-al \cite{12} was just a mathematical jugglery which went absolutely in vain. At this end we would like to add that the authors \cite{12} used the point Lagrangian $L(a, R, \dot a, \dot R)= 6a \dot a^2 F' + 6a^2\dot a\dot R F'' - a^3(F' R - F)$, with a wrong sign. It is interesting to notice that the Noether equations remain unaltered even for such toppling of signs between the kinetic and the potential terms.

\subsection{Conserved currents provided in \cite{13}:}

Apart from the general symmetry associated with arbitrary form of $F(R)$, admitting  trivial first integral in the form of energy, which is constraint to vanish in gravitational theory, Paliathanasis et-al \cite{13} obtained a host of different forms of $F(R)$ along with the conserved currents, under the request of Noether symmetries of $F(R)$ theory of gravity in matter dominated era. In this subsection our aim is to analyze the symmetries so obtained.\\

\noindent
\textbf{Case-I:}\\

\noindent
It has been claimed that Noether symmetry is admissible for $F(R) = F_0 R^{\frac{3}{2}}$ in flat space ($k = 0$), and is associated with the following three different conserved currents in the presence of cold dark matter, listed below.

\be I_2 =\frac{d}{dt} ({a\sqrt R}),\ee
\be I_3 =t \frac{d}{dt} ({a\sqrt R})-{a\sqrt R},\ee
\be\label{sigma2} I_4 =  2t E-6a^2\dot a \sqrt R-6a^3\frac{\dot R}{\sqrt R}= 2tE-6a^2\bigg(\dot a \sqrt R+\frac{a\dot R}{\sqrt R}\bigg).\ee
We have already analysed $I_2$, which is $\Sigma_1$ \eqref{sigma}, and $I_3$, where, ${dI_3\over dt} = t {d^2 \Sigma_1\over dt^2}$ is just a modified version of $\Sigma_1$ \eqref{sigma}, and therefore are new ones. It is therefore only required to analyse $I_4$. The Hamiltonian and its derivatives are already presented in \eqref{H-1} and \eqref{HD1} respectively. The expression of conserved current $I_4$ in terms of generalized coordinates and generalized momenta may be found in view of \eqref{GC1} as,

\be I_4 = 2t \left[\frac{4 R^{\frac{3}{2}} P^2_R}{9F_0 a^3}-\frac{2 R^{\frac{1}{2}} P_a P_R}{9 F_0a^2}+\frac{1}{2}F_0a^3 R^{\frac{3}{2}}+\rho_{0}\right]+\frac{4a P_a}{3F_0}-{4\over F_0}R P_R =2t H +\frac{4a P_a}{3F_0}-{4\over F_0}R P_R, \ee
whose derivatives are the following,

\be\begin{split}&
\frac{\partial {I_{4}}}{\partial {a}}=2t\frac{\partial {H}}{\partial {a}} +\frac{4 P_a}{3F_0},
\hspace{1.14 in}\frac{\partial {I_{4}}}{\partial {P_a}}= 2t\frac{\partial {H}}{\partial {P_a}}+\frac{4a}{3F_0},\\&
\frac{\partial {I_{4}}}{\partial {R}}= 2t\frac{\partial {H}}{\partial {R}}-{4\over F_0} P_R,
\hspace{1.02 in}\frac{\partial {I_{4}}}{\partial {P_R}}= 2t\frac{\partial {H}}{\partial {P_R}}-{4R\over F_0},\hspace{0.8 in}
\frac{\partial {I_{4}}}{\partial {t}}=  2H.
\end{split}\ee
Thus one finds,
\be \frac {d{I_{4}}}{dt} = [{I_{4}},H]+\frac{\partial {I_{4}}}{\partial {t}}={2\over F_0}\left[\frac{4R^\frac{3}{2} P_{R}^2}{9F_0 a^3} - \frac{2R^\frac{1}{2}P_{a}P_{R}}{9 F_0a^2} + \frac{F_0}{2} a^3R^\frac{3}{2}\right]+2H=2H\Big(1 + {1\over F_0}\Big)-{2\rho_{0}\over F_0} = 0.\ee
It vanishes due to diffeomorphic invariance in vacuum ($\rho_{0}= 0$), \textbf{but not in the matter dominated era}, as claimed erroneously in \cite{13}. Since for gravity, the energy is constrained to vanish \eqref{E}, therefore the above conserved current $I_4$ associated with $F(R) \propto R^{3\over 2}$ may be expressed in the following  compact manner,

\be\label{sigma3} I_4 =  a^2\bigg(\dot a \sqrt R+\frac{a\dot R}{\sqrt R}\bigg),\ee
for $k = 0$, in vacuum. Note that, while the conserved current $\Sigma_1$ \eqref{sigma} appears from the first pair of field equations \eqref{FE1}, this conserved current $I_4$ \eqref{sigma3} appears from the combination of all the three \eqref{FE1}.\\

\noindent
\textbf{Case-II:}\\

\noindent
The authors \cite{13} also claimed that in flat space ($k = 0$), Noether symmetry exists even for $F(R) = F_0R^{\frac{7}{8}}$, admitting two different conserved currents in the presence of cold dark matter, which are,
\be \label{3.2I5} I_5 =2tE-\frac{21}{8}\frac{d}{dt} ({a^3 R^{-\frac{1}{8}}})=2tE -\frac{63}{8} a^2 \dot a R^{-\frac{1}{8}}+\frac{21}{64} a^3 R^{-\frac{9}{8}} \dot R ,\ee
\be \label{3.2I6} I_6 =t^2 E- \frac{21}{8}t \frac{d}{dt} ({a^3 R^{-\frac{1}{8}}})+\frac{21}{8}{a^3 R^{-\frac{1}{8}}}=t^2 E -\frac{63}{8} t a^2 \dot a R^{-\frac{1}{8}}+\frac{21}{64} t a^3 R^{-\frac{9}{8}} \dot R +\frac{21}{8} a^3 R^{-\frac{1}{8}}.\ee
The Hamiltonian and the phase-space structures of the conserved currents may be found as before in the forms,

\be \label{H56} H =\frac{32R^\frac{9}{8}P_{a}P_{R}}{21F_0 a^2}+\frac{256R^\frac{17}{8} P_{R}^2}{21 F_0a^3}-\frac{F_0}{8} a^3 R^\frac{7}{8}+\rho_{0}.\ee
\be\label{I5} {I_{5}}=2t\left[\frac{32R^\frac{9}{8}P_{a}P_{R}}{21 F_0a^2}+\frac{256R^\frac{17}{8} P_{R}^2}{21F_0 a^3}-\frac{F_0}{8} a^3 R^\frac{7}{8}+\rho_{0}\right]+ \frac{a P_{a}}{2 F_0} -{4R P_{R}\over F_0} = 2tH + \frac{a P_{a}}{2 F_0} -{4R P_{R}\over F_0},\ee
\be \begin{split}\label{I6}{I_{6}}&= t^2\left[\frac{32R^\frac{9}{8}P_{a}P_{R}}{21F_0 a^2}+\frac{256R^\frac{17}{8} P_{R}^2}{21 F_0a^3}-\frac{F_0}{8} a^3 R^\frac{7}{8}+\rho_{0}\right] + t\Big(\frac{a P_{a}}{2F_0} -{4R P_{R}\over F_0}\Big)+\frac{21a^3 }{8R^{\frac{1}{8}}}\\&
= t^2 H + t\Big(\frac{a P_{a}}{2F_0} -{4R P_{R}\over F_0}\Big)+\frac{21a^3 }{8R^{\frac{1}{8}}}.\end{split}\ee
The partial derivatives of the Hamiltonian \eqref{H56} are,

\be\label{HI5}\begin{split}&
\frac{\partial {H}}{\partial {a}}=- \frac{64R^\frac{9}{8}P_{a}P_{R}}{21F_0 a^3}-\frac{256R^\frac{17}{8} P_{R}^2}{7 F_0a^4}-\frac{3}{8} F_0a^2 R^\frac{7}{8},\hspace{0.3 in}\frac{\partial {H}}{\partial {P_a}}=\frac{32R^\frac{9}{8}P_{R}}{21F_0 a^2},\\&
\frac{\partial {H}}{\partial {R}}= \frac{12R^\frac{1}{8}P_{a}P_{R}}{7F_0 a^2}+\frac{544R^\frac{9}{8} P_{R}^2}{21F_0 a^3}-\frac{7}{64}F_0 a^3 R^{-\frac{1}{8}},\hspace{0.29 in}
\frac{\partial {H}}{\partial {P_R}}=\frac{32R^\frac{9}{8}P_{a}}{21F_0 a^2}+\frac{512R^\frac{17}{8} P_{R}}{21 F_0a^3}.\end{split}\ee

\noindent
\textbf{Subcase-I:}\\

\noindent
Partial derivatives of $I_5$ \eqref{I5} are,
\be\label{I55}\begin{split}&
\frac{\partial {I_{5}}}{\partial {a}}=2t\frac{\partial {H}}{\partial {a}} + \frac{P_{a}}{2F_0} ,\hspace{1.3 in}\frac{\partial {I_{5}}}{\partial {P_a}}=2t\frac{\partial {H}}{\partial {P_a}} + \frac{a}{2F_0},\\&
\frac{\partial {I_{5}}}{\partial {R}}=2t\frac{\partial {H}}{\partial {R}} - {4 P_{R}\over F_0},\hspace{1.26 in}
\frac{\partial {I_{5}}}{\partial {P_R}}= 2t\frac{\partial {H}}{\partial {P_R}}-{4 R\over F_0},\hspace{0.55 in}
\frac{\partial {I_{5}}}{\partial t} = 2H.
\end{split}\ee
Computing the time derivative of the conserved current, one therefore arrives at,

\be \begin{split}&\frac {d{I_{5}}}{dt} = [{I_{5}},H]+ \frac{\partial {I_{5}}}{\partial t} = {2\over F_0}\left[\frac{32 R^\frac{9}{8} P_{a} P_{R}}{21F_0 a^2}+\frac{128  R^\frac{17}{8} P_{R}^2}{21 F_0a^3}-\frac{F_0}{8} a^3 R^\frac{7}{8}\right]+2H \\&= {2\over F_0}(H-\rho_0) + 2H =2H\left(1+ {1\over F_0}\right)-{2\rho_{0}\over F_0}.\end{split}\ee
Since $\frac {d{I_{5}}}{dt} = 0$ due to the diffeomorphic invariance $H = 0$, in the absence of matter $\rho_0 = 0$, therefore, Noether symmetry exists for $F(R) \propto R^{7\over 8}$, and $I_5$ is indeed the associated conserved current, but only in vacuum, \textbf{and not in the presence of matter}, as erroneously claimed by the authors \cite{13}.\\

\noindent
\textbf{Subcase-II:}\\

\noindent
Here, we work with the same Hamiltonian \eqref{H56}, whose partial derivatives are already presented in \eqref{HI5}. We therefore present the derivatives of the conserved current $I_6$ given in \eqref{I6} as,
\be\begin{split}&
\frac{\partial {I_{6}}}{\partial {a}}=t^2\frac{\partial {H}}{\partial {a}} + t\frac{P_{a}}{2F_0}+\frac{63 a^2}{8 R^{\frac{1}{8}}},
\hspace{0.4 in}
\frac{\partial {I_{6}}}{\partial {P_a}}=t^2\frac{\partial {H}}{\partial {P_a}}+ t\frac{ a}{2F_0} ,\\&
\frac{\partial {I_{6}}}{\partial {R}}=t^2\frac{\partial {H}}{\partial {R}} - t{4 P_{R}\over F_0}-\frac{21}{64} {a^3\over R^{\frac{9}{8}}},
\hspace{0.25 in}\frac{\partial {I_{6}}}{\partial {P_R}}= t^2\frac{\partial {H}}{\partial {P_R}} - t{4 R\over F_0},\hspace{0.3 in}
\frac{\partial {I_{6}}}{\partial t} =  2Ht + \frac{a P_{a}}{2F_0} -{4R P_{R}\over F_0}.
\end{split}\ee
One can thus calculate the the time derivative of the conserved current as,
\be \frac {d{I_{6}}}{dt} = [{I_{6}},H]+ \frac{\partial {I_{6}}}{\partial t} = {2t\over F_0}\left[\frac{32 R^\frac{9}{8} P_{a} P_{R}}{21F_0 a^2}+\frac{256  R^\frac{17}{8} P_{R}^2}{21F_0 a^3}-\frac{F_0}{8}  a^3 R^\frac{7}{8}\right]+2t H = 2t\left[H\Big(1 + {1\over F_0}\Big) - {\rho_0\over F_0}\right] = 0.\ee
Above expression vanishes again due to diffeomorphic invariance for $\rho_0 =0$, which confirms that $I_6$ is also the conserved current associated with the Hamiltonian for $F(R) \propto R^{7\over 8}$, in the vacuum, \textbf{and not in the matter dominated era}. However, one can easily observe that ${dI_6\over dt} = I_5$, and thus the two \eqref{3.2I5} and \eqref{3.2I6} are the same conserved currents in disguise. Thus, one can finally assert that $F(R) \propto R^{7\over 8}$ admits a conserved current,

\be\label{sigma4} I_5 = \frac{d}{dt} ({a^3 R^{-\frac{1}{8}}}),\ee
in vacuum, for ($k = 0$), since as mentioned $E = 0$, in view of \eqref{E}. Nevertheless, this symmetry ($F(R) \propto R^{7\over 8}$) implies $F'' < 0$, i.e. the final attractor is not a de Sitter point.\\

\noindent
\textbf{Case-IV:}\\

\noindent
Finally, the authors \cite{13} also could explore Noether symmetry for $F(R) \propto R^{n}$, in flat space ($k = 0$), where $n$ is arbitrary other than, $n = {3\over 2}$ and $n = {7\over 8}$, carrying the following conserved current,

\be\label{I7}{I_{7}}= 2tE-8n(2-n) \dot{a} a^2 {R}^{n-1}-4n(n-1)(2n-1) a^3 \dot{R}R^{n-2}\ee
In view of the above form of $F(R) = F_0 R^{n}$, the Hamiltonian \eqref{H1}, and using equations \eqref{GC1}, the phase-space structure of the above conserved current ${I_7} $ may be expressed as,

\be\begin{split} &H =-\frac{P_{a}P_{R}}{6n(n-1)F_0  a^2R^{n-2}}+\frac{P_{R}^2}{6n (n-1)^2F_0 a^3R^{n-3} } +(n-1)F_0  a^3R^n+ \rho_{0},\\&{I_{7}}  =2t\left[-\frac{P_{a}P_{R}}{6n(n-1)F_0  a^2R^{n-2}}+\frac{P_{R}^2}{6n (n-1)^2F_0  a^3R^{n-3} } +(n-1)F_0  a^3R^n+ \rho_{0}\right] + \frac{2(2n-1)a P_{a}}{3 F_0 }-{4R P_{R}\over F_0},\\&
~~~=2t H + \frac{2(2n-1)a P_{a}}{3 F_0 }-{4R P_{R}\over F_0}.
\end{split}\ee

One can therefore compute,
\be\begin{split}&
\frac{\partial {H}}{\partial {a}} = \frac{P_a P_R}{3n(n-1)F_0 a^3 R^{n-2}}-\frac{P_R^2}{2n(n-1)^2 F_0a^4 R^{n-3}}+3(n-1)F_0a^2 R^{n},\hspace{0.15 in}\frac{\partial {H}}{\partial {P_a}}= -\frac{P_R}{6n(n-1)F_0a^2 R^{n-2}},\\&
\frac{\partial {H}}{\partial {R}} = -\frac{(2-n)P_a P_R}{6n(n-1)F_0 a^2 R^{n-1}}+\frac{(3-n) P_R^2}{6 n(n-1)^2F_0 a^3 R^{n-2}}+n(n-1)F_0a^3 R^{n-1},\\&\frac{\partial {H}}{\partial {P_R}}=-\frac{P_a}{6n(n-1)F_0a^2 R^{n-2}} +\frac{P_R}{3n(n-1)^2F_0 a^3 R^{n-3}},\\&
\frac{\partial {I_{7}}}{\partial {a}} = 2t\frac{\partial {H}}{\partial {a}} +\frac{2(2n-1)P_a}{3F_0},\hspace{0.6 in}\frac{\partial {I_{7}}}{\partial {P_a}}=2t\frac{\partial {H}}{\partial {P_a}}+ \frac{2(2n-1)a}{3F_0},\\&
\frac{\partial {I_{7}}}{\partial {R}} =2t\frac{\partial {H}}{\partial {P_a}} -{4P_R \over F_0},\hspace{1.04 in}\frac{\partial {I_{7}}}{\partial {P_R}}=2t\frac{\partial {H}}{\partial {P_R}} - {4R\over F_0},\hspace{0.7 in}\frac{\partial {I_{7}}}{\partial {t}} = 2 H.
\end{split}\ee

\noindent
and in view of the above relations, one can at once check that,

\be \begin{split}\frac {d{I_{7}}}{dt} &= [{I_{7}},H]+\frac{\partial {I_{6}}}{\partial t}= {2\over F_0}\left[\frac{P_{R}^2}{6n (n-1)^2 a^3R^{n-3}} -\frac{P_{a}P_{R}}{6n(n-1) a^2R^{n-2}} + (n-1) a^3R^n \right]+2H \\&= {2\over F_0}(H-\rho_{0}) + 2H = 2H \left(1+{1\over F_0}\right) - {2\rho_0\over F_0},\end{split}\ee
which again vanishes due to diffeomorphic invariance in the vacuum dominated era ($\rho_0 = 0$), confirming that $F(R)$ indeed admits a symmetry for $F(R) \propto R^n$, carrying the conserved current $I_{7}$ \eqref{I7} in vacuum and again \textbf{not in the matter dominated era}. However, it is not clear, why the authors suggested $n \ne {3\over 2}$ and $n \ne {7\over 8}$. One can clearly generate the conserved currents \eqref{sigma3} for $n = {3\over 2}$ and \eqref{sigma4} for $n = {7\over 8}$. Note that for $n = {7\over 8}$, $F'' < 0$, and so the final attractor is not a de Sitter point. However, one can also recover the conserved current $\Sigma_2 = a^3\dot R$ \eqref{R2} for $n = 2$, i.e. for $F(R) \propto R^2$, as well. It may be mentioned that this conserved current for $F(R) \propto R^2$ exists for arbitrary $k$ \eqref{R2}.\\

\noindent
Thus, the only new finding in \cite{13} is: In flat space ($k = 0$), $F(R) \propto R^n$ is an outcome of Noether symmetry analysis in vacuum dominated era, for which the conserved current is,

\be\label{sigma5}I= 2(2-n) \dot{a} a^2 {R}^{n-1} + (n-1)(2n-1)) a^3 \dot{R}R^{n-2},\ee
where we have set $E =0$, since energy is constrained to vanish. Unfortunately, it does not recover the conserved currents associated with $F(R) \propto R^{-1}$ \eqref{-1} and $F(R) \propto R^{7\over 5}$ \eqref{75}, which means at least a pair of conserved currents is available for both, as in the case of $F(R) \propto R^{3\over 2}$. Although, all the cases studied and the associated results obtained by the authors \cite{13} have been summed up to a single result, it is important to note that all the symmetries expatiated here separately satisfy the field equations. The reason being, the authors involved the constraint $E = 0$ in the symmetry analysis.

\subsection{Conserved currents provided in \cite{14}:}

It is already stated that $F(R)$ theory of gravity might admit additional symmetries under different choice of $\eta$ in \eqref{gen}, which were not explored, since our aim in \cite{7} was to establish equivalence between different frames. Applying Killing tensors of the minisuperspace, in which the field equations are invariant under contact transformations, Paliathanasis \cite{14} obtained several forms of $F(R)$, taking perfect fluid in the form of pressure-less dust (correspondingly for vacuum too) in the flat space ($k = 0$). The symmetries were explored in Jordan's frame and then translated to find the forms of $F(R)$ using the transformation laws, $\phi=F'(R)$ and the first order Clairaut equation $V(\phi)=(F' R-F)$. In the following we analyze all the symmetries obtained in \cite{14} case-by-case.\\

\noindent
\textbf{Case-I:}\\

\noindent
The first form of $F(R)$ and the associated conserved charge obtained in \cite{14} in the flat space ($k = 0$) are,
\be \label{I1} F(R) = F_0 (R-V_{1})^{\frac{3}{2}},~{I_{1}} = 3(\dot{a}\phi+a\dot{\phi})^2 -V_{1}a^2 {\phi}^2= {27F_0^2\over 16(R-V_1)}\bigg[\Big(a\dot R + 2\dot{a}(R-V_{1})\Big)^2 - {4\over 3}V_{1}a^2 (R-V_{1})^2\bigg],\ee
where, we have used the relations $\phi = F', ~\dot\phi = F''\dot R$, required to establish scalar-tensor equivalence of $F(R)$ theory of gravity.
The corresponding Hamiltonian for the above form of $F(R)$ \eqref{I1} may be expressed in view of \eqref{H1}, and the phase-space structure of the conserved current ${I_{1}}$ appearing in \eqref{I1} may be cast in view of \eqref{GC1} as,
\be \begin{split} &H_{1}=\frac{4(R-V_1)^\frac{3}{2}}{9 F_0 a^3}P_{R}^2-\frac{2(R-V_1)^\frac{1}{2}}{9 F_0 a^2}P_{a}P_{R}+ F_0 a^3 (R-V_1)^\frac{1}{2} \Big(\frac{R}{2}+V_1\Big)+\rho_0,\\&{I_{1}}=\frac{P_{a}^2}{12a^2}+\frac{(R-V_1)^2}{3a^4} P_{R}^2-\frac{(R-V_1)}{3a^3} P_{a} P_{R}-\frac{9 F_0^2a^2}{4} V_1(R-V_1).\end{split}\ee

As a result, one can compute,
\be\begin{split}&
\frac{\partial {H_{1}}}{\partial {a}}={(R-V_1)^{1\over 2}}\Big[\frac{4P_{a}P_{R}}{9 F_0 a^3}-\frac{4(R-V_1) P_{R}^2}{3 F_0 a^4} +3 F_0 a^2 \Big(\frac{R}{2}+V_1\Big)\Big],\hspace{0.1 in}\frac{\partial {H_{1}}}{\partial {P_a}}= -\frac{2(R-V_1)^\frac{1}{2} P_R}{9 F_0 a^2},\\&
\frac{\partial {H_{1}}}{\partial {R}}= \frac{1}{(R-V_1)^{\frac{1}{2}}}\Big[\frac{2(R-V_1) P_{R}^2}{3 F_0 a^3}-\frac{P_{a}P_{R}}{9 F_0 a^2} + \frac{3F_0}{4}a^3 R\Big],\hspace{0.05 in}\frac{\partial {H_{1}}}{\partial {P_R}}= \frac{8(R-V_1)^\frac{3}{2} P_R}{9 F_0 a^3}-\frac{2(R-V_1)^\frac{1}{2} P_a}{9 F_0 a^2},\\&
\frac{\partial {I_{1}}}{\partial {a}}= -\frac{P^2_a}{6a^3}-(R-V_1)\Big[\frac{4(R-V_1) P^2_R}{3a^5}-\frac{P_{a} P_{R}}{a^4}+\frac{9aV_1 F^2_0}{2}\Big]
,\hspace{0.27 in}\frac{\partial {I_{1}}}{\partial {P_a}}= \frac{P_a}{6a^2}-\frac{(R-V_1) P_{R}}{3a^3}\\&\frac{\partial {I_{1}}}{\partial {R}}= \frac{2(R-V_1) P^2_R}{3a^4}-\frac{P_{a} P_{R}}{3a^3}-\frac{9 a^2 V_1 F^2_0}{4},\hspace{0.83 in}\frac{\partial {I_{1}}}{\partial {P_R}}= \frac{2(R-V_1)^2 P_R}{3a^4}-\frac{(R-V_1) P_{a}}{3a^3}.
\end{split}\ee
In view of the above relations, one can at once check that
\be \begin{split}\frac {d{I_{1}}}{dt} = 0.\end{split}\ee
Since $I_1$ vanishes identically, being independent of $\rho_0$, so it is indeed the conserved current associated with the Hamiltonian $H_1$, which applies both in the vacuum era $\rho_0 = 0$, and in the matter dominated era $\rho_0 = \rho_{m0}$. It is definitely interesting to learn that $F(R)$ theory admits a symmetry for $F(R) \propto (R - V_1)^{3\over 2}$ which generalizes the commonly known one $F(R) \propto R^{3\over 2}$, in the presence of cosmological constant $\Lambda = V_1$. This was also noticed earlier by the authors \cite{13}. As the Poisson bracket vanishes trivially, so it is a symmetry that supposed to have found for $\eta = 0$ in equation \eqref{gen}, but it was not. However, one can take note of the fact that setting $V_1$ to vanish, the conserved current $I_1 \propto \frac{1}{R}(a\dot R + 2\dot a R)^2$ is essentially the square of the known standard one $\Sigma_1 \propto {1\over \sqrt R}(a\dot R + 2\dot a R)$, appearing in Case-1 \eqref{sigma}, of subsection (3.1). \\

\noindent
As mentioned in the introduction, if for a function of phase-space variables, Poisson bracket vanishes identically in the absence of explicit time dependence, then the function is an integral of motion of the theory under consideration, and therefore must satisfy the field equations. In fact the current form of $F(R)$ and the associated conserved current \eqref{I1} is also valid for arbitrary $k$. For the sake of demonstration, we write the field equations in view of the point Lagrangian \eqref{L1}, and for the present form of the potential $V(\phi)=V_1 \phi + V_2 \phi^3$, for $k = 0, \pm 1$, as,

\be\begin{split}\label{1.1} &2\frac{\ddot{a}}{a}+\frac{\dot{a}^2}{a^2} + {k\over a^2}  +\frac{\ddot{\phi}}{\phi}+2\frac{\dot{a}\dot{\phi}}{a\phi}-\frac{V_1}{2}
-\frac{V_2 \phi^2}{2}=0,\\&
\frac{\ddot{a}}{a}+\frac{\dot{a}^2}{a^2} + {k\over a^2}-\frac{V_1}{6}-\frac{ V_2 \phi^2}{2}=0,\\&
\frac{\dot{a}^2}{a^2} + {k\over a^2} +\frac{\dot{a}\dot{\phi}}{a\phi}-\frac{V_1\phi^3}{6} - \frac{V_2\phi^5}{6} - \frac{\rho_{0}}{6a^3\phi}=0.\end{split}\ee
Now, the difference of the first pair of equations \eqref{1.1} yields,

\be \label{Comb1}\left(\frac{\ddot{a}}{a}+2\frac{\dot{a}\dot{\phi}}{a\phi}
+\frac{\ddot{\phi}}{\phi}\right) - \frac{V_1}{3} = 0.\ee
One can note at a glance that the time derivative of conserved current \eqref{I1},

\be\frac {d{I_{1}}}{dt}=2a\phi(\dot{a}\phi+a\dot{\phi})\left[3\left(\frac{\ddot{a}}{a}+2\frac{\dot{a}\dot{\phi}}{a\phi}
+\frac{\ddot{\phi}}{\phi}\right)-V_1\right],\ee
vanishes due to the above equation \eqref{Comb1}, independent of the presence of the energy constraint equation and the matter density, and also for arbitrary $k = 0, \pm 1$.\\

\noindent
\textbf{Case-II:}\\

\noindent
The second form of $F(R)$ and the associated conserved charge obtained are,
\be \begin{split} \label{7by8} &F(R) = F_0(R-V_{1})^{\frac{7}{8}},\\&{I_{2}} = 3a^4(\phi \dot{a}-a\dot{\phi})^2+4V_{2}a^6 {\phi}^{-6} \\&~~~~= \frac{147F^2_0 a^4}{64}\left(\frac{\dot{a}^2}{(R-V_{1})^\frac{1}{4}}+ \frac{ a^2\dot{R}^2}{64(R-V_{1})^\frac{9}{4}}+\frac{a\dot{a}\dot{R}}{4(R-V_{1})^\frac{5}{4}}\right)
+4 \Big(\frac{8}{7F_0}\Big)^{6} V_{2}a^6(R-V_{1})^\frac{3}{4}.\end{split}\ee
This again generalizes the previous case appearing in case-II of subsection (3.3), and presented in \cite{13} earlier. As before the Hamiltonian, the phase-space structure of the conserved current ${I_{2}}$ may be expressed in views of in view of \eqref{H1} and \eqref{GC1} respectively as,

\be \begin{split} &H_{2} = \frac{32(R-V_1)^\frac{9}{8}P_{a}P_{R}}{21F_0 a^2} + \frac{256(R-V_1)^\frac{17}{8} P_{R}^2}{21F_0 a^3} - {F_0 a^3(R-8V_1)\over 8(R-V_1)^\frac{1}{8}} + \rho_0,\\&{I_{2}}=\frac{a^2P_{a}^2}{12}+48(R-V_1)^2 P_{R}^2+{4a(R-V_1)P_a P_R}+4 \Big(\frac{8a}{7F_0}\Big)^{6} V_{2}(R-V_{1})^\frac{3}{4}.\end{split}\ee
As usual, one can therefore compute

\be\begin{split}&
\frac{\partial {H_{2}}}{\partial {a}}= -\frac{64(R-V_1)^\frac{9}{8}P_{a}P_{R}}{21F_0 a^3} - \frac{256(R-V_1)^\frac{17}{8} P_{R}^2}{7 F_0 a^4} - {3 F_0 a^2(R-8V_1)\over 8(R-V_1)^\frac{1}{8}},~\frac{\partial {H_{2}}}{\partial {P_a}}= \frac{32(R-V_1)^\frac{9}{8}P_{R}}{21F_0 a^2},\\&
\frac{\partial {H_{2}}}{\partial {R}}= \frac{12(R-V_1)^\frac{1}{8}P_{a}P_{R}}{7F_0 a^2}+\frac{544(R-V_1)^\frac{9}{8} P_{R}^2}{21F_0 a^3} - \frac{7F_0 a^3R}{64(R-V_1)^\frac{9}{8}},~\frac{\partial {H_{2}}}{\partial {P_R}}= \frac{32(R-V_1)^\frac{9}{8}}{21F_0 a^2}\Big[P_{a}+\frac{16(R-V_1)^\frac{1}{8}P_{R}}{a}\Big],\\&
\frac{\partial {I_{2}}}{\partial {a}}= \frac{a P_{a}^2}{6}+{4(R-V_1)P_{a} P_{R}}+24\Big(\frac{8}{7F_0}\Big)^{6} V_{2}a^5(R-V_{1})^\frac{3}{4},~~\frac{\partial {I_{2}}}{\partial {P_a}}= \frac{a^2 P_a}{6}+{4 a (R-V_1) P_{R}},\\&
\frac{\partial {I_{2}}}{\partial {R}}= 96(R-V_1) P_{R}^2+{4 a P_{a} P_{R}}+3 \Big(\frac{8}{7F_0}\Big)^{6} {V_{2}a^6\over(R-V_{1})^{\frac{1}{4}}},~~~~~~\frac{\partial {I_{2}}}{\partial {P_R}}= 96(R-V_1)^2 P_{R}+{4a(R-V_1)P_a}.
\end{split}\ee
In view of the above relations, one can again check that
\be\begin{split} \frac {d{I_{2}}}{dt} &= {F_0\over 2}a^3 (R-V_1)^{7\over 8}\left[1 + 7\Big({8\over 7F_0}\Big)^8V_2\right]\Big(aP_a + 24(R-V_1)P_R\Big) = 0,\end{split}\ee
provided $V_2 = -{1\over 7}\big({7F_0\over 8}\big)^8$ \footnote{Note that, the author erroneously claimed a different relation \big($V_2 = {1\over 7}\big({7F_0\over 8}\big)^{8\over 7}$\big) between $F_0$ and $V_2$.}. It was therefore claimed that $I_2$ \eqref{7by8} is the conserved current associated with the Hamiltonian $H_2$, both in the vacuum era $\rho_0 = 0$, as well as in the matter dominated eras $\rho_0 = \rho_{m0}$. However, as stated, if for a function of phase-space variables, the first theorem of Poisson is satisfied conditionally, then the function is not an integral of motion. In the appendix we shall prove that in view of the field equations the time derivative of the above conserved current seizes to vanish.\\

\noindent
\textbf{Case-III:}\\

\noindent
The third form of $F(R)$ and the associated conserved charge obtained are,
\be \label{1by3} F(R) = F_0 R^{\frac{1}{3}} - V_{1},~~~{I_{3}} = 6a^3\dot{a}(a\dot{\phi}-\phi\dot{a})-a^5\left(\frac{3V_{1}}{5}-V_{2}{\phi}^{-\frac{1}{2}}\right)= -\frac{2F_0 a^3\dot{a}^2}{R^\frac{2}{3}} -\frac{4F_0 a^4\dot{a}\dot{R}}{3R^\frac{5}{3}}-\frac{3V_1 a^5}{5}+\sqrt{\frac{3}{F_0}}V_2 R^\frac{1}{3}.\ee
Now, the Hamiltonian \eqref{H1} and the phase-space structure of the conserved current ${I_{3}}$ in view of equation \eqref{GC1}, are expressed as,
\be\begin{split}& H_{3}=\frac{3 R^\frac{5}{3} P_{a}P_{R}}{4F_0 a^2}+\frac{9R^\frac{8}{3} P_{R}^2}{8F_0 a^3}- \frac{2F_0 a^3 R^\frac{1}{3}}{3} +V_1 a^3+\rho_0,\\&{I_{3}} = -\frac{27R^\frac{8}{3}P_R^2}{8F_0 a}-\frac{3R^\frac{5}{3}P_a P_R}{4F_0}-\frac{3V_1 a^5}{5}+\sqrt{\frac{3}{F_0}}V_2 a^5 R^\frac{1}{3},\end{split}\ee
and thus, one can compute,

\be\begin{split}&
\frac{\partial {H_{3}}}{\partial {a}}= -\frac{3R^\frac{5}{3} P_{a}P_{R}}{2F_0 a^3} - \frac{27R^\frac{8}{3} P_{R}^2}{8F_0 a^4} - {2 F_0 a^2 R^\frac{1}{3}} + {3 V_1 a^2},
\hspace{0.3in}\frac{\partial {H_{3}}}{\partial {P_a}}=\frac{3R^\frac{5}{3}P_{R}}{4F_0 a^2}\\&
\frac{\partial {H_{3}}}{\partial {R}}= \frac{5R^\frac{2}{3} P_{a}P_{R}}{4F_0 a^2} + \frac{3R^\frac{5}{3} P_{R}^2}{F_0 a^3} -\frac{2F_0 a^3}{9R^\frac{2}{3}},
\hspace{1.14 in}\frac{\partial {H_{3}}}{\partial {P_R}}= \frac{3R^\frac{5}{3} P_{a}}{4F_0 a^2}+\frac{9R^\frac{8}{3} P_{R}}{4F_0 a^3}\\&
\frac{\partial {{I_{3}}}}{\partial {a}}= \frac{27R^\frac{8}{3}P_R^2}{8F_0 a^2}-3V_1 a^4+{5\sqrt{\frac{3}{F_0}}V_2 a^4 R^\frac{1}{3}},
\hspace{0.94 in}\frac{\partial {{{I_{3}}}}}{\partial {P_a}}=-\frac{3R^\frac{5}{3}P_{R}}{4F_0},\\&
\frac{\partial {{{I_{3}}}}}{\partial {R}}= -\frac{5R^\frac{2}{3}P_a P_R}{4F_0}-\frac{9R^\frac{5}{3}P_R^2}{F_0 a}+\frac{\sqrt{\frac{3}{F_0}}V_2 a^5 R^\frac{-2}{3}}{3},
\hspace{0.6 in}\frac{\partial {{{I_{3}}}}}{\partial {P_R}}= -\frac{3R^\frac{5}{3}P_a}{4F_0}-\frac{27R^\frac{8}{3}P_R}{4F_0 a}
\end{split}\ee
In view of the above relations, one can at once check that
\be\begin{split} \frac {d{I_{3}}}{dt}& = \left(3a^2 R^2 P_{R}+\frac{a^3 R P_{a}}{6}\right)\left[{3\over 2F_0}\sqrt{3\over F_0}V_2 -1\right] = 0,\end{split}\ee
under the condition $ V_2 = 2\left({F_0\over 3}\right)^{3\over 2}$, as indicated in \cite{14}. Since the relation is independent of $\rho_0$ so, it was claimed that $I_3$ \eqref{1by3} is indeed the conserved current associated with the Hamiltonian $H_3$, in the vacuum era $\rho_0 = 0$ as well as in the matter dominated era $\rho_0 = \rho_{m0}$. Again we mention that, since derivative of $I_3$ vanishes conditionally, so it is not an integral of motion for the $F(R) = F_0 R^{\frac{1}{3}} - V_{1}$ form. As for the previous case, we shall again prove this fact in the appendix from the point of view of the field equations.\\

\noindent
\textbf{Case-IV:}\\

\noindent
The authors further enlisted in case IV of the article \cite{14}, three additional forms of $F(R)$ along with their associated conserved charges, which we take up here one-by-one.\\

\noindent
\textbf{Case-IVa:}\\

\noindent
The first of these are,

\be \label{CC} F(R) = F_{0}R^{\frac{3}{2}},~~~{I^1_{4}} = 12a^2(a^2\dot{\phi}^2 - \phi^2 \dot{a}^2)+3V_{1}(a \phi)^4=\frac{27 F^2_0 a^4 \dot{R}^2}{4R}-27 F^2_0 a^2 \dot{a}^2 R + \frac{243 V_1 F^4_0 a^4 R^2}{16}.\ee
As before the Hamiltonian and phase-space structure of the conserved current ${I^1_{4}}$ take the following forms,

\be\begin{split}& H^1_{4}=\frac{4R^\frac{3}{2} P_{R}^2}{9F_{0} a^3}-\frac{2R^\frac{1}{2}P_{a}P_{R}}{9 F_{0} a^2} + \frac{F_{0}a^3 R^\frac{3}{2}}{2} + \rho_0,\\& {I^1_{4}}=\frac{P_{a}^2}{3} +\frac{4 {R}^2 P_{R}^2}{{a}^2}-\frac{8R P_{a} P_{R} }{3a}+\frac{243 V_1 F^4_0 a^4 R^2}{16},\end{split}\ee
and so one finds,

\be\begin{split}&
\frac{\partial {H^1_{4}}}{\partial {a}} = \frac{4R^\frac{1}{2}P_a P_R}{9F_{0}a^3} -\frac{4R^\frac{3}{2} P_R^2}{3F_{0}a^4} + \frac{3F_{0}a^2R^\frac{3}{2}}{2},\hspace{0.6 in}
\frac{\partial {H^1_{4}}}{\partial {P_a}} = -\frac{2R^\frac{1}{2} P_R}{9F_{0}a^2}.\\&
\frac{\partial {H^1_{4}}}{\partial {R}} = -\frac{R^{-\frac{1}{2}}P_a P_R}{9F_{0}a^2}+\frac{2R^\frac{1}{2} P_R^2}{3F_{0}a^3} + \frac{3F_{0}a^3R^\frac{1}{2}}{4},\hspace{0.48 in}\frac{\partial {H^1_{4}}}{\partial {P_R}} = -\frac{2R^\frac{1}{2} P_a}{9F_{0}a^2} + \frac{8R^\frac{3}{2} P_R}{9F_{0}a^3}.\\&
\frac{\partial {I^1_{4}}}{\partial {a}} = \frac{8 R P_a P_R}{3a^2}-\frac{8R^2 P^2_R}{a^3}+\frac{243 V_1 F^4_0 a^3 R^2}{4},\hspace{0.49 in}\frac{\partial {I^1_{4}}}{\partial {P_a}} = \frac{2P_a}{3}-\frac{8RP_R}{3a}.\\&
\frac{\partial {I^1_{4}}}{\partial {R}} = -\frac{8 P_a P_R}{3a}+\frac{8R P^2_R}{a^2}+\frac{243 V_1 F^4_0 a^4 R}{8},\hspace{0.615 in}\frac{\partial {I^1_{4}}}{\partial {P_R}} = -\frac{8RP_a}{3a}+\frac{8R^2P_R}{a^2}.
\end{split}\ee
In view of the above relations, one can at once check that

\be \begin{split}\frac {d{I^1_{4}}}{dt}& = [{I^1_{4}},H^1_{4}]= {F_0\over 4}a R^{3\over 2}(2RP_R - a P_a)(27 F_0^2 V_1 - 4) = 0,\end{split}\ee
provided, $V_1 = {4\over 27F_0^2}$, as indicated in \cite{14}, and it is again independent of $\rho_{0}$. Thus, the natural claim was: $I^1_{4}$ is indeed the conserved current associated with the Hamiltonian $H^1_{4}$, both in the vacuum era $\rho_0 = 0$ as well as in the matter dominated era $\rho_0 = \rho_{m0}$. As already mentioned, such claim is erroneous, since $I^1_{4}$ vanishes conditionally, as we shall again prove in the appendix, .\\

\noindent
\textbf{Case-IV b:}\\

\noindent
The second form of $F(R)$ and the associated conserved charge obtained are,

\be\label{R4} F(R) = F_{0}R^{4},\hspace{0.1 in}{I^2_{4}} = 12a^2(a^2\dot{\phi}^2 - \phi^2 \dot{a}^2)+4V_{2}a^4 {\phi}^5=1728 F^2_0 a^4 R^4 \dot{R}^2-192 F^2_0 a^2 \dot{a}^2 R^6 + 4096 V_2 F^5_0 a^4 {R}^{15}. \ee
The  Hamiltonian and the conserved current ${I^1_{4}}$ in phase-space variables, are expressed as,

\be\begin{split}& H^2_{4}=\frac{P_{R}^2}{216 F_{0} a^3 R}-\frac{P_{a}P_{R}}{72F_{0} a^2 R^2} + {3F_{0}a^3 R^4} + \rho_0\\& {I^2_{4}}=\frac{P_{a}^2}{3} +\frac{R^2 P_{R}^2}{9{a}^2}-\frac{4R P_{a} P_{R}}{9a}+4096 V_2 F^5_0 a^4 {R}^{15},\end{split}\ee

and so, one finds,
\be\begin{split}&
\frac{\partial {H^2_{4}}}{\partial {a}}= \frac{P_{a}P_{R}}{36F_{0} a^3 R^2} - \frac{P_{R}^2}{72 F_{0} a^4 R} + {9F_{0}a^2 R^4},\hspace{0.7 in}\frac{\partial {H^2_{4}}}{\partial {P_a}}= -\frac{P_{R}}{72F_{0} a^2 R^2},\\&
\frac{\partial {H^2_{4}}}{\partial {R}}=\frac{P_{a}P_{R}}{36F_{0} a^2 R^3}-\frac{P_{R}^2}{216 F_{0} a^3 R^2} + {12F_{0}a^3 R^3},
\hspace{0.5 in}\frac{\partial {H^2_{4}}}{\partial {P_R}}= -\frac{P_{a}}{72 F_{0} a^2 R^2}+\frac{P_{R}}{108F_{0} a^3 R}\\&
\frac{\partial {I^2_{4}}}{\partial {a}}= -\frac{2R^2 P_{R}^2}{9{a}^3}+\frac{4R P_{a} P_{R}}{9a^2}+16384 V_2 F^5_0 a^3 {R}^{15},\hspace{0.35 in}\frac{\partial {I^2_{4}}}{\partial {P_a}}= \frac{2P_a}{3}-\frac{4RP_R}{9a},\\&
\frac{\partial {I^2_{4}}}{\partial {R}}= \frac{2 R P_{R}^2}{9{a}^2}-\frac{4 P_{a} P_{R} }{9a}+61440 V_2 F^5_0 a^4 {R}^{14},
\hspace{0.62 in}\frac{\partial {I^2_{4}}}{\partial {P_R}}= \frac{2R^2 P_R}{9a^2} - \frac{4 R P_a}{9a}.
\end{split}\ee
In view of the above relations, one can at once check that
\be \begin{split}\frac {d{I^2_{4}}}{dt} &= [{I^2_{4}},H^2_{4}]= \frac{4}{3}F_0 a R^5 P_R \left[1+256 V_2 F^3_0 R^8\right] - \frac{2}{3}F_0 a^2 R^4 P_a \left[1 + 1280 V_2 F^3_0 R^8\right]\neq 0.\end{split}\ee
It was claimed in \cite{14} that a symmetry exists for the form of $F(R)$ and the conserved current given above in \eqref{R4} under the condition $F_0 = {3\over 8}\big({2\over V_2}\big)^{1\over 3}$. However, one can clearly observe that the Poisson bracket does not vanish, which confirms that $I^2_{4}$ is not the conserved current associated with the Hamiltonian $H^2_{4}$. \textbf{Therefore $F(R)$ theory of gravity does not admit a symmetry associated with $F(R) = F_{0}R^{4}$}, and so the claim is faulty.\\

\noindent
\textbf{Case-IV c:}\\

\noindent
The third form of $F(R)$, together with the associated conserved charge and the form of potential obtained are,
\be \label{comb} F(R) = \frac{V_1 R}{4V_2}+\frac{3V_1R^\frac{2}{3}}{8V_2 g^\frac{1}{3}}+\frac{3g^\frac{1}{3}R^\frac{4}{3}}{16V_2 },~~~{I^3_{4}} = 12a^2(a^2\dot{\phi}^2 - \phi^2 \dot{a}^2)+ 3V_{1}(a \phi)^4 + 4V_{2}a^4 {\phi}^5,~~~V({\phi})=(V_1 \phi^3 + V_2 \phi^4), \ee

\noindent
where, $V_1$, $V_2$ are constants. Since $F(R)$ containing three terms, therefore to avoid unnecessary complication we compute the Poisson bracket in Jordan's frame. The expression of  Lagrangian in the flat space ($k = 0$), spanned by a set of configuration space variable $(a, \phi, \dot a, \dot {\phi})$ in Jordan's frame \cite{6, 7} is presented in \eqref{L1}. Note that, in the vacuum era ($\rho_0 = 0$), while for the pressure-less dust era ($\rho_{0} = \rho_{m0}$), so that one can compute the Poisson bracket for both the cases together. The Hamiltonian \eqref{H2} together with the phase-space structure of the conserved current ${I^3_{4}}$ may now be expressed as,

\be \begin{split}\label{HI3} & H^3_{4}(a, \phi, P_{a}, P_{\phi}) =  -\frac{P_{a}P_{\phi}}{6a^2} + \frac{\phi P_{\phi}^2}{6a^3} + V_1 a^3\phi^3 + V_2 a^3\phi^4 + \rho_0 =0,\\&
{I^3_{4}}=\frac{P_{a}^2}{3} +\frac{{\phi}^2 P_{\phi}^2}{{a}^2}-\frac{4\phi P_{a} P_{\phi} }{3a}+3V_{1}a^4 \phi^4 + 4V_{2}a^4 {\phi}^5.\end{split}\ee

As a result, one can compute,
\be\begin{split}&
\frac{\partial {H^3_{4}}}{\partial {a}}= \frac{P_{a}P_{\phi}}{3 a^3} - \frac{\phi P_{\phi}^2}{2a^4}+ 3a^2(V_1 \phi^3 + V_2 \phi^4), \hspace{1.0 in}\frac{\partial {H^3_{4}}}{\partial {P_a}}= -\frac{P_{\phi}}{6 a^2 },\\&\frac{\partial {H^3_{4}}}{\partial {\phi}}= \frac{P_{\phi}^2}{6a^3} + a^3(3 V_1 {\phi}^2 + 4 V_2 \phi^3),\hspace{1.49 in}\frac{\partial {H^3_{4}}}{\partial {P_{\phi}}}= -\frac{P_{a}}{6 a^2} + \frac{\phi P_{\phi}}{3 a^3}\\&
\frac{\partial {I^3_{4}}}{\partial {a}}= -\frac{2{\phi}^2 P_{\phi}^2}{{a}^3}+\frac{4\phi P_{a} P_{\phi}}{3a^2}+12 V_{1}a^3 \phi^4 + 16 V_{2}a^3 {\phi}^5,\hspace{0.42 in}\frac{\partial {I^3_{4}}}{\partial {P_a}}= \frac{2P_a}{3}-\frac{4\phi P_{\phi}}{3a},\\&
\frac{\partial {I^3_{4}}}{\partial {\phi}}= \frac{2\phi P_{\phi}^2}{{a}^2}-\frac{4 P_{a} P_{\phi}}{3a}+12V_{1}a^4 \phi^3 + 20 V_{2}a^4 {\phi}^4,
\hspace{0.67 in}\frac{\partial {I^3_{4}}}{\partial {P_{\phi}}}= - \frac{4 \phi P_a}{3a}+\frac{2{\phi}^2 P_{\phi}}{a^2}.
\end{split}\ee
In view of the above relations, one can use \eqref{P2} to check that,
\be \begin{split}\frac {d{I^3_{4}}}{dt} &= [{I^3_{4}},H^3_{4}]= \left(\frac{\partial {I^3_{4}}}{\partial a}\frac{\partial H^3_{4}}{\partial P_{a}}- \frac{\partial {I^3_{4}}}{\partial P_{a}}\frac{\partial {H^3_{4}}}{\partial a}\right) + \left(\frac{\partial {I^3_{4}}}{\partial \phi}\frac{\partial H^3_{4}}{\partial P_{\phi}}- \frac{\partial {I^3_{4}}}{\partial P_{\phi}}\frac{\partial {H^3_{4}}}{\partial \phi}\right)=  0,\end{split}\ee
which confirms that $I^3_{4}$ is indeed the conserved current associated with the Hamiltonian $H^3_{4}$, and Noether symmetry admits a combination of three terms in $F(R)$ including the linear term, which is admissible both in the vacuum and in matter dominated era. This is highly encouraging, since it might be possible to find a Friedmann-like early deceleration followed by late-time acceleration.\\

\noindent
\textbf{Case-V:}\\

\noindent
The last form of $F(R)$ and the associated conserved charge obtained are,
\be\label{VII1}\begin{split} &F(R) = F_{0}R^\frac{4}{3}-\frac {V_1 R}{4 V_2},~~~{I_{5}} = 12a^2[(\beta-\phi^2)\dot{a}^2+a^2\dot{\phi}^2]-a^4(\beta-\phi^2)[V_{1}(\beta+3\phi^2)+4V_{2}(3\beta\phi+\phi^3)]=\\&  \bigg[12a^2 \dot a^2(-\frac{16 F^2_0 R^\frac{2}{3}}{9}+\frac{2V_1 F_0 R^\frac{1}{3}}{3V_2})+\frac{64F^2_0 a^4 \dot R^2}{27 R^\frac{4}{3}}\bigg]+\frac{4096 F^5_0 R^\frac{5}{3} V_2 a^4}{243}-\frac{512F^4_0 R^\frac{4}{3} V_1 a^4}{81}\end{split}\ee

\noindent
The corresponding Hamiltonian for the above form of $F(R)$ \eqref{VII1} may be expressed in view of \eqref{H1}, and the phase-space structure of the conserved current ${I_{5}}$ appearing in \eqref{VII1} may be cast in view of \eqref{GC1} as,
\be\begin{split} &H_{5}=-\frac{3R^\frac{2}{3}P_{a}P_{R}}{8F_{0} a^2}+\frac{9R^\frac{5}{3}P^2_{R}}{8F_{0} a^3}-\frac{27V_1 R^\frac{4}{3}P^2_{R}}{128V_2 F^2_{0} a^3}+\frac{F_0 a^3 R^\frac{4}{3}}{3}+\rho_{0},\\&{I_{5}}=\frac{P_{a}^2}{3} +{9R^2\over a^2}\left(1-
\frac{3V_1 R^{-{1\over 3}}}{8V_2 F_0} +\frac{3V^2_1 R^{-{2\over 3}}}{64V^2_2 F^2_0}\right)P^2_{R}-\frac{4R}{a}\left(1 - \frac{3V_1 R^{-\frac{2}{3}}}{16V_2 F_0}\right)P_{a} P_{R}+\frac{512F^4_0 a^4 {R}^\frac{4}{3}}{81}\left(\frac{8 V_2 F_0 {R}^\frac{1}{3}}{3}-V_1\right).\end{split}\ee
As a result, one can compute,
\be\begin{split}&
\frac{\partial {H_{5}}}{\partial {a}}= \bigg[\frac{81V_1 R^\frac{4}{3}}{128V_2 F^2_{0} a^4}-\frac{27R^\frac{5}{3}}{8F_{0} a^4}\bigg]P^2_{R}+\frac{3R^\frac{2}{3}P_{a}P_{R}}{4F_{0} a^3}+ F_0 a^2 R^\frac{4}{3},\hspace{0.06 in}
\frac{\partial {H_{5}}}{\partial {P_a}}= -\frac{3R^\frac{2}{3}P_{R}}{8F_{0} a^2}\\&
\frac{\partial {H_{5}}}{\partial {R}}=\bigg[\frac{15R^\frac{2}{3}}{8F_{0} a^3}-\frac{9V_1 R^\frac{1}{3}}{32V_2 F^2_{0} a^3}\bigg] P^2_{R} -\frac{P_{a}P_{R}}{4F_{0} a^2R^{\frac{1}{3}}}+\frac{4F_0 a^3 R^\frac{1}{3}}{9},\hspace{0.05 in}
\frac{\partial {H_{5}}}{\partial {P_R}}= \frac{9R^\frac{5}{3}P_{R}}{4F_{0} a^3}-\frac{27V_1 R^\frac{4}{3}P_{R}}{64V_2 F^2_{0} a^3}-\frac{3R^\frac{2}{3} P_{a}}{8F_{0} a^2}\\&
\frac{\partial {I_{5}}}{\partial {a}}= \bigg[\frac{27V_1 R^\frac{5}{3}}{4V_2 F_0 a^3}-\frac{18R^2}{a^3}-\frac{27V^2_1 R^\frac{4}{3}}{32V^2_2 F^2_0 a^3}\bigg]P^2_{R}+\bigg[\frac{4R}{a^2}-\frac{3V_1 R^\frac{2}{3}}{4V_2 F_0 a^2}\bigg] P_{a} P_{R}\\&\hspace{0.4 in}+\frac{16384 V_2 F^5_0 a^3 {R}^\frac{5}{3}}{243}-\frac{2048V_1 F^4_0 a^3 {R}^\frac{4}{3}}{81},\hspace{0.84 in}\frac{\partial {I_{5}}}{\partial {P_a}}= \frac{2P_a}{3}-\frac{4RP_R}{a}+\frac{3V_1 R^\frac{2}{3} P_R}{4V_2 F_0 a}\\&
\frac{\partial {I_{5}}}{\partial {R}}= \bigg[\frac{18R}{a^2}-\frac{135V_1 R^\frac{2}{3}}{24V_2 F_0 a^2}+\frac{9V^2_1 R^\frac{1}{3}}{16V^2_2 F^2_0 a^2}\bigg]P^2_{R}+\bigg[\frac{V_1 }{2V_2 F_0 aR^{\frac{1}{3}}}-\frac{4 }{a}\bigg]P_{a} P_{R}+{2048\over729}(10 V_2 F_0{R}^\frac{2}{3}-3V_1 {R}^\frac{1}{3})F^4_0 a^4 ,\\&
\frac{\partial {I_{5}}}{\partial {P_R}}= \bigg[\frac{18R^2}{a^2}-\frac{27V_1 R^\frac{5}{3}}{4V_2 F_0 a^2}+\frac{27V^2_1 R^\frac{4}{3}}{32V^2_2 F^2_0 a^2}\bigg]P_{R}+\bigg[\frac{3V_1 R^\frac{2}{3}}{4V_2 F_0 a}-\frac{4R }{a}\bigg]P_{a}\end{split}\ee
In view of the above relations, one can at once check that
\be \begin{split}\frac {d{I_{5}}}{dt} &= [{I_{5}},H_{5}]+\frac{\partial {I_{5}}}{\partial {t}}\\&=
\bigg[\frac{V_1a^2 R P_a}{3V_2}-\frac{10F_0 a R^\frac{4}{3}P_a}{9}+4F_0 a R^\frac{7}{3}P_{R}-\frac{9V_1 a R^2 P_R}{4V_2}+\frac{3V^2_1 a R^\frac{5}{3}P_R}{8V^2_2 F_0}\bigg]\bigg(\frac{256V_2 F^3_0}{27} - 1\bigg)= 0,\end{split}\ee
under the condition, $F_{0}=(\frac{27}{256 V_2})^\frac{1}{3}$, which does not match with the the one presented in \cite{14}. However, as pointed out earlier that such conditional vanishing of the time derivative of $I_{5}$ confirms that $I_{5}$ is not the conserved current associated with the Hamiltonian $H_{5}$, and therefore such symmetry is obscure, as we shall explore in the appendix.\\

\noindent
Therefore, at the end, the overall outcome of the work \cite{14} are the existence of Noether symmetries for $F(R) \propto (R - V_1)^{3\over 2}$, and $F(R) = \alpha R + \beta R^{2\over 3} + \gamma R^{4\over 3}$ for which the conserved currents are presented in \eqref{I1} and \eqref{comb} respectively. Rest of the symmetries seize to exist, since as repeatedly mentioned, neither the energy constraint had not been taken into account in the symmetry analysis, nor the field equations had been checked.

\section{Constructing a generalized action in view of available symmetry.}

In view of the Poisson first theorem we have proved that not all the symmetries explored over years by different authors, following Noether symmetry analysis, are the relevant symmetries of the theory under consideration. In this section, we present a list of available (viable) symmetries of $F(R)$ theory of gravity, appearing in the literature till date for future reference.\\

\noindent
\textbf{Vacuum era.}\\

\noindent
1. $F(R) = F_0 (R-V_{1})^{\frac{3}{2}}.$\\
\be {Q_{1}}= {1\over (R-V_1)}\bigg[\Big(a\dot R + 2\dot{a}(R-V_{1})\Big)^2 - {4\over 3}V_{1}a^2(R-V_{1})^2\bigg],~~~\mathrm{for}~k=0, \pm 1.\ee

\noindent
2. $F(R)=F_0 R^n.$
\be{Q_{2}}= 2(2-n) \dot{a} a^2 {R}^{n-1}+(n-1)(2n-1)) a^3 \dot{R}R^{n-2},~~~\mathrm{for}~k=0.\ee
However, for $F(R)=F_0 R^2$ in particular, the conserved current ($\Sigma = a^3\dot R$) holds for ($k = 0, \pm 1$).\\

\noindent
3. $F(R) =\frac{ F_0}{R}.$
\be {Q_{3}}= R{\sqrt a}~\dot a,~~~\mathrm{for}~k=0.\ee

\noindent
4. $ F_{0}R^{\frac{7}{5}},$
\be {Q_{4}}=  \sqrt a {d\over dt}(a R^{2\over 5}),~~~\mathrm{for}~k=0.\ee

\noindent
5. $F(R) = \frac{V_1 R}{4V_2}+\frac{3V_1R^\frac{2}{3}}{8V_2 g^\frac{1}{3}}+\frac{3g^\frac{1}{3}R^\frac{4}{3}}{16V_2}.$
\be{Q_{5}}= 12a^2(a^2\dot{\phi}^2 - \phi^2 \dot{a}^2)+ 3V_{1}(a \phi)^4 + 4V_{2}a^4 {\phi}^5,~~~\mathrm{for}~k=0.\ee

\noindent
\textbf{Radiation era.}\\

\noindent
1. $F(R) = F_0 R^2$
\be{Q_{6}}  =  a^3\dot R,~~~\mathrm{for}~k=0, \pm 1.\ee

\noindent
\textbf{Pressureless dust era.}\\

\noindent
1. $F(R) = F_0 (R-V_{1})^{\frac{3}{2}}.$
\be {Q_{7}}= {1\over (R-V_1)}\bigg[\Big(a\dot R + 2\dot{a}(R-V_{1})\Big)^2 - {4\over 3}V_{1}a^2(R-V_{1})^2\bigg],~~~\mathrm{for}~k=0, \pm 1.\ee

\noindent
2. $F(R) = \frac{V_1 R}{4V_2}+\frac{3V_1R^\frac{2}{3}}{8V_2 g^\frac{1}{3}}+\frac{3g^\frac{1}{3}R^\frac{4}{3}}{16V_2 }.$
\be{Q_{8}}= 12a^2(a^2\dot{\phi}^2 - \phi^2 \dot{a}^2)+ 3V_{1}(a \phi)^4 + 4V_{2}a^4 {\phi}^5,~~~\mathrm{for}~k=0.\ee

\noindent
What to do now with all these symmetries is definitely a viable question to ask. The answer is: one can now construct, for example, an action in the following form,

\be \label{Action} A = \int\left[\alpha R + \beta R^2 + \gamma R^{4\over3} + \delta  R^{2\over 3} + \mathcal{L}_m\right]\sqrt{-g} ~d^4 x,\ee
(say), to study early universe till date, where, $\mathcal{L}_m$ stand for matter Lagrangian. One can note that in the early universe, $R^2$ gives the dominant contribution leading to Inflation, while in the late universe, it may be neglected, and in view of the symmetry associated with rest of the terms, one might be able to expatiate early Friedmann-like deceleration, with late stage of cosmic acceleration. The reason being, in the middle, linear term ($\alpha R$) would contribute the most, while at the end, $R^{2\over 3}$ term would dominate, which might lead to late-time accelerated expansion. Other suitable combinations are also possible.

\section{Concluding remarks.}

In a nut-shell, gravity being associated with diffeomorphic invariance, resulting in Hamiltonian and three momenta constraints, Noether symmetry is not on-shell unless, the constraints are incorporated through Lagrange multipliers in the symmetry generator. For the situation where momenta constraints are absent (in the absence of space-time components in the metric), one should use equation \eqref{gen}. Since it is required to go through a huge number of trial and errors for selecting appropriate $\eta$ in search of symmetry, one can still use the standard symmetry generator. However, in that case, the constraint equations are not supposed to be satisfied by the symmetry obtained following Noether symmetry analysis, since the generator has no given input regarding the fact that the theory admits constraints. Thus, it is necessary to check if the conserved current satisfies the constraint equations, in particular. Our present finding that some of the conserved currents appearing in the literature, are not viable symmetries of the theory under consideration, is due to the fact that the authors restrained themselves from such a check. Nonetheless, it often turns out to be a difficult or at least very tedious job. In this article we propose an easier alternative of using the first theorem of Poisson \eqref{P1}. However, the statement for obtaining an integral of motion for the theory under consideration, is modified. In connection with the Hamiltonian constraint, it now states that ``\emph{If the right hand side of the total derivative of a phase-space variable vanishes identically, or is a function of the Hamiltonian (which is constrained to vanish), the phase space variable is an integral of motion, otherwise it is not, even if it vanishes conditionally}". The reason is: the Hamiltonian constraint equation does not admit such condition. The technique is simple and elegant to prove if the conserved currents so obtained following symmetry analysis, are indeed the integrals of motion of the theory under consideration.\\

Under Noether symmetry analysis, different authors applied different techniques to obtain a host of conserved currents associated with different forms of $F(R)$, in pure $F(R)$ theory of gravity. The present study reveals the fact that some of these so-called conserved currents are not the integral of motions of the theory under consideration. For example, our findings are:\\
1. Out of the four conserved currents associated with the same form of $F(R) \propto R^{3\over 2}$ provided by Shamir et-al in \cite{12}, the first two are faulty, the fourth is well-known, viz. $\Sigma_1 = {d\over dt} (a\sqrt R)$, and the third is the same $\Sigma_1$, in disguise. Thus, this article \cite{12} does not produce any new result, despite considering an additional time derivative of the boundary term.\\
2. Paliathanasis et-al \cite{13} on the contrary presented a host of forms of $F(R)$ associated with respective conserved currents, and claimed that all these hold in matter dominated era. Present analysis reveals that the only new result emerges out of this analysis is $F(R) \propto R^n$, where $n$ is arbitrary, which holds only in vacuum era and not in the matter dominated era.\\
3. Finally, Paliathanasis \cite{14} have also made a systematic study independently, and presented a host of conserved currents associated with different forms of $F(R)$. Present analysis reveals that the forms $F(R) \propto (R- V_1)^{3\over 2}$ and $F(R) = \alpha R + \beta R^{2\over 3} + \gamma R^{4\over 3}$ indeed result in symmetry, in vacuum as well as in the matter dominated eras. On the contrary, the claims $F(R) \propto (R - V_1)^{7\over 8}$, $F(R) = F_0 R^{1\over 3} - V_1$, $F(R) \propto R^4$ and $F(R) = -\alpha R + \beta R^{4\over 3}$ and a different conserved current associated with $F(R) \propto R^{3\over 2}$ are faulty, since the conserved currents obtained vanish only conditionally, which do not satisfy the Hamilton constraint equation. This we expatiate in the appendix.\\

We have briefly discussed the importance of the available forms of $F(R)$ following symmetry analysis. Note that most of the forms are admissible in vacuum dominated era, and therefore should be studied in the context of very early universe. Further, it is well understood that in the matter dominated era, the universe must have evolved with early decelerated expansion, followed by recent accelerated expansion. Such a puzzle can only be solved by a combination of scalar invariant terms appearing in the action. In this sense, the form $F(R) = \alpha R + \beta R^{2\over 3} + \gamma R^{4\over 3}$ obtained in \cite{14} is encouraging. It would be interesting to study the evolution of the universe at the early stage (in the context of inflation), the radiation era in the middle, and the matter dominated era; in view of the proposed action \eqref{A}. This we pose in future.

\appendix
\section{Analysing cases in which Poisson Theorem satisfied conditionally:}

We have stated that in the case of gravity, in which the Hamiltonian ($H$) is constrained to vanish, if the right hand side of the first Poison theorem ${dI\over dt} = [I, H] + {\partial I \over \partial t}$, for a function of phase-space variables $I = I (q_i, p_i, t)$, vanishes trivially, or turns out to be proportional to the Hamiltonian $H$, then $I$ is an integral of motion of the system under consideration. However, we have come across at least four cases, where the authors claimed to have obtained Noether symmetries, in which $dI\over dt$ vanishes conditionally. These situations appear in case II, III, IVa and in V of the article by Paliathanasis \cite{14}, which we have studied in section (3.4). In view of the field equations, we justify that such claims are erroneous, .\\

\noindent
\textbf{Section 3.4, Case II:}\\

\noindent
The form of $F(R)$, the associated conserved charge and the form of potential obtained are,
\be F(R) = F_{0}(R-V_1)^{\frac{7}{8}},~~~{I_{2}} =  3a^4(\dot{a}\phi-a\dot{\phi})^2 +4V_{2}a^6 {\phi}^{-6},~~~V(\phi)=V_1 \phi - V_2 \phi^{-7},~~~F_0 = {8\over 7}(7V_2)^{7\over 8}. \ee
Now, the expression of time derivative of conserved current is
\be\frac {d{I_{2}}}{dt}=6a^5 \phi(\dot{a}\phi-a\dot{\phi})\left[\left(\frac{\ddot{a}}{a}+2\frac{\dot{a}^2}{a^2}-2\frac{\dot{a}\dot{\phi}}{a\phi}
-\frac{\ddot{\phi}}{\phi}\right)+4V_2 \phi^{-8}\right].\ee
\noindent
For the above potential, the field equations \eqref{FE2} read as
\be\begin{split}& 2\frac{\ddot{a}}{a}+\frac{\dot{a}^2}{a^2}+\frac{\ddot{\phi}}{\phi}+2\frac{\dot{a}\dot{\phi}}{a\phi}-\frac{V_1}{2}
+\frac{V_2 \phi^{-8}}{2}=0,\\&
\frac{\ddot{a}}{a}+\frac{\dot{a}^2}{a^2}-\frac{V_1}{6}-\frac{7 V_2 \phi^{-8}}{6}=0,\\&
\frac{\dot{a}^2}{a^2}+\frac{\dot{a}\dot{\phi}}{a\phi}-\frac{V}{6\phi}-\frac{\rho_{0}}{6a^3\phi}=0.\end{split}\ee

\noindent
Eliminating $V_1$ term between the first pair of field equations, one arrives at,

\be\left(\frac{\ddot{a}}{a}+2\frac{\dot{a}^2}{a^2}-2\frac{\dot{a}\dot{\phi}}{a\phi}
-\frac{\ddot{\phi}}{\phi}\right)=4V_2 \phi^{-8},\ee
which makes apparent that $\frac{d{I_{2}}}{dt} \ne 0$.
To make things more convincing, one can notice that the conserved current expressed in terms of ($a,R,\dot a, \dot R$) \eqref{I1} does not reduce to either the conserved current $I_5$ or $I_6$, appearing in section (3.3) associated with $F = F_0 R^{7\over 8}$, setting $V_1 = 0$.\\

\noindent
\textbf{Section 3.4, Case III:}\\

\noindent
The form of $F(R)$, the associated conserved charge and the form of potential obtained are,
\be F(R) = F_0 R^{\frac{1}{3}} - V_{1},~~~{I_{3}} =
6a^3\dot{a}(a\dot{\phi}-\phi\dot{a})-a^5\left(\frac{3V_{1}}{5}-{V_{2}\over \sqrt\phi}\right),~~~V(\phi)= V_1 -\frac{V_2}{ \phi^\frac{1}{2}},~~~ F_0 = 3\left({V_2\over 2}\right)^{2\over 3}.\ee
We take the time derivative of $I_{3}$ as,
\be\label{PN}\frac{d{I_{3}}}{dt}=6a^5 \dot{\phi}\left[\frac{\ddot{a}}{a}-2\frac{\ddot{a}}{a}\frac{\dot{a}}{a}\frac{\phi}{\dot{\phi}}+3\frac{\dot{a}^2}{a^2}
-3\frac{\dot{a}^3}{a^3}+\frac{\dot{a}}{a}\frac{\ddot{\phi}}{\dot{\phi}}\right]-3V_1 a^5\Big(\frac{\dot{a}}{a}\Big)
+{a^5 V_2\over\sqrt\phi} \left(5\frac{\dot{a}}{a}-\frac{\dot{\phi}}{2\phi}\right) \ee
For the above potential, the field equations \eqref{FE2} read as,

\be\begin{split}& 2\frac{\ddot{a}}{a}+\frac{\dot{a}^2}{a^2}+\frac{\ddot{\phi}}{\phi}+2\frac{\dot{a}\dot{\phi}}{a\phi}-\frac{V_1}{2\phi}
+\frac{V_2}{\phi^\frac{3}{2}}=0,\\&
\frac{\ddot{a}}{a}+\frac{\dot{a}^2}{a^2}-\frac{V_2}{12\phi^\frac{3}{2}}=0,\\&
\frac{\dot{a}^2}{a^2}+\frac{\dot{a}\dot{\phi}}{a\phi}-\frac{V_1}{6\phi} + \frac{V_2}{6\phi^{3\over 2}}-\frac{\rho_{0}}{6a^3\phi}=0.\end{split}\ee

Now, we find the values of $V_1$ and $V_2$ from the first pair of field equations
\be\begin{split}&
V_1=2\phi\left[14\frac{\ddot{a}}{a}+13\frac{\dot{a}^2}{a^2}+\frac{\ddot{\phi}}{\phi}+2\frac{\dot{a}\dot{\phi}}{a\phi}\right],\\&
V_2=12\phi^\frac{3}{2}\left[\frac{\ddot{a}}{a}+\frac{\dot{a}^2}{a^2}\right].\end{split}\ee
Substituting the values of $V_1$ and $V_2$ in equation \eqref{PN}, we obtained

\be\begin{split}&
\frac {d{I_{3}}}{dt}=6a^5 \dot{\phi}\left[\frac{\ddot{a}}{a}-2\frac{\ddot{a}}{a}\frac{\dot{a}}{a}\frac{\phi}{\dot{\phi}}+3\frac{\dot{a}^2}{a^2}
-3\frac{\dot{a}^3}{a^3}+\frac{\dot{a}}{a}\frac{\ddot{\phi}}{\dot{\phi}}\right]-6a^4\dot{a} \phi \left[14\frac{\ddot{a}}{a}+13\frac{\dot{a}^2}{a^2}+\frac{\ddot{\phi}}{\phi}+2\frac{\dot{a}\dot{\phi}}{a\phi}\right]   +12a^5\phi\left[\frac{\ddot{a}}{a}+\frac{\dot{a}^2}{a^2}\right]\left[5\frac{\dot{a}}{a}-\frac{\dot{\phi}}{2\phi}\right]
,\end{split}\ee
which clearly implies, $\frac {d{I_{3}}}{dt}\ne0$\\

\noindent
\textbf{Section 3.4, Case IVa:}\\

\noindent
The form of $F(R)$, the associated conserved charge and the form of potential obtained are,
\be F(R) = F_{0}R^{\frac{3}{2}},~~~{I^1_{4}} = 12a^2(a^2\dot{\phi}^2 - \phi^2 \dot{a}^2) + 3V_{1}a^4 {\phi}^4,~~~V(\phi)= V_1 \phi^3. \ee
As before, we take the time derivative of conserved current, which reads as,

\be \label{I4}\frac {d{I^1_{4}}}{dt}=-12a^3 \phi^2 \dot{a}\left[2\frac{\ddot{a}}{a}+2\frac{\dot{a}^2}{a^2}-2\frac{a \dot{\phi}\ddot{\phi}}{\dot{a}\phi^2}-4\frac{\dot{\phi}^2}{\phi^2}+2\frac{\dot{a}\dot{\phi}}{a\phi}-V_1 {\phi^2}-\frac{V_1 a\phi \dot{\phi}}{\dot{a}}\right],\ee
\noindent
For the above potential, the field equations \eqref{FE2} take the form,

\be \begin{split} &2\frac{\ddot{a}}{a}+\frac{\dot{a}^2}{a^2}+\frac{\ddot{\phi}}{\phi}+2\frac{\dot{a}\dot{\phi}}{a\phi}-\frac{V_1 \phi^2}{2}=0,\\&
\frac{\ddot{a}}{a}+\frac{\dot{a}^2}{a^2}-\frac{ V_1 \phi^2}{2}=0,\\&
\frac{\dot{a}^2}{a^2}+\frac{\dot{a}\dot{\phi}}{a\phi}-\frac{V_1\phi^2}{6}-\frac{\rho_{0}}{6a^3\phi}=0.\end{split}\ee

\noindent
Multiplying the first of the field equations by a factor $(-2\frac{a\dot{\phi}}{\dot{a}\phi})$, and the second by a factor $2$ and under addition, one arrives at,

\be 2\frac{\ddot{a}}{a}+2\frac{\dot{a}^2}{a^2}-2\frac{a \dot{\phi}\ddot{\phi}}{\dot{a}\phi^2}-4\frac{\dot{\phi}^2}{\phi^2}-V_1 {\phi^2}-2\frac{\dot{a}\dot{\phi}}{a\phi}+\frac{V_1 a\phi \dot{\phi}}{\dot{a}}-4\frac{\ddot{a}\dot{\phi}}{\dot{a}\phi} = 0,\ee
which is clearly different from \eqref{I4}. Thus $I^1_{4}$ is not an integral of motion.\\

\noindent
\textbf{Section 3.4, Case V:}\\

\noindent
The forms of $F(R)$ and the conserved current $I_5$ in flat space ($k = 0$) are ,

\be \begin{split} &F(R) = F_{0}R^\frac{4}{3}-\frac {V_1 R}{4 V_2},\\&
{I_5}=\bigg[12a^2 \dot a^2\left(-\frac{16 F^2_0 R^\frac{2}{3}}{9}+\frac{2V_1 F_0 R^\frac{1}{3}}{3V_2}\right)+\frac{64F^2_0 a^4 \dot R^2}{27 R^\frac{4}{3}}\bigg]+\frac{4096 F^5_0 R 6\frac{5}{3} V_2 a^4}{243}-\frac{512F^4_0 R^\frac{4}{3} V_1 a^4}{81}.\end{split}\ee
The field equations \eqref{FE1} are the following,

\be\begin{split}& 2\frac{\ddot{a}}{a}+\frac{\dot{a}^2}{a^2}-\frac{3 V_1 \dot{a}^2}{16 V_2 F_0 a^2 R^{\frac{1}{3}}}-\frac{3 V_1 \ddot{a}}{8 V_2 F_0 a R^{\frac{1}{3}}}+\frac{2\dot{a}\dot{R}}{3aR}+\frac{\ddot{R}}{3R}-\frac{2\dot{R}^2}{9R^2}+\frac{R}{8}=0,\\&
\frac{\ddot{a}}{a}+\frac{\dot{a}^2}{a^2}+\frac{R}{18}=0,\\&
\frac{\dot{a}^2}{a^2}-\frac{3 V_1 \dot{a}^2}{16 V_2 F_0 a^2 R^{\frac{1}{3}}}-\frac{\dot{a}\dot{R}}{3aR}+\frac{R}{24}=0.\end{split}\ee

\noindent
Taking the time derivative of conserved current $I_5$, we find,

\be\label{PO}\begin{split}\frac {d I_5}{dt}=&\frac{\ddot{a}}{a}+\frac{\dot{a}^2}{a^2}+\frac{\dot{a}\dot{R}}{3aR}+\frac{3V_1\dot{a}^2}{8V_2 F_0 a^2 R^{\frac{1}{3}}}+\frac{3V_1\ddot{a}}{8V_2 F_0 a R^{\frac{1}{3}}}+\frac{V_1\dot{a}\dot{R}}{16V_2 F_0 a R^{\frac{4}{3}}}+\frac{2F^2_0\dot{R}^2}{9R^2}\\&
+\frac{2F^2_0 a\dot{R}\ddot{R}}{9R^2}-\frac{2F^2_0 a\dot{R}^3}{27R^{\frac{7}{3}}}+\frac{128V_2 F^3_0}{81 R^{-1}}+\frac{160V_2 F^3_0 a\dot{R}}{243}-\frac{16V_1 F^2_0}{27 R^{-\frac{2}{3}}}-\frac{16V_1 F^2_0 a\dot{R}}{81 R^{\frac{1}{3}}}=0.\end{split}\ee
Now, arranging the first and the third field equations we find,
\be -2\frac{\ddot{a}}{a}-2\frac{\dot{a}^2}{a^2}+\frac{3 V_1 \dot{a}^2}{8 V_2 F_0 a^2 R^{\frac{1}{3}}}+\frac{3 V_1 \ddot{a}}{8 V_2 F_0 a R^{\frac{1}{3}}}-\frac{\dot{a}\dot{R}}{3aR}-\frac{\ddot{R}}{3R}+\frac{2\dot{R}^2}{9R^2}-\frac{R}{6}=0,\ee

Multiplying the second field equation by a factor 3 and under addition with the above equation,
one arrives at,
\be \frac{\ddot{a}}{a}+\frac{\dot{a}^2}{a^2}+\frac{3 V_1 \dot{a}^2}{8 V_2 F_0 a^2 R^{\frac{1}{3}}}+\frac{3 V_1 \ddot{a}}{8 V_2 F_0 a R^{\frac{1}{3}}}-\frac{\dot{a}\dot{R}}{3aR}-\frac{\ddot{R}}{3R}+\frac{2\dot{R}^2}{9R^2}-\frac{R}{9}=0,\ee
which does not match with the time derivative of $I_5$ \eqref{PO}. It is therefore apparent that conserved current does not satisfy the field equation.\\

\end{document}